# A two-phase mixing layer between parallel gas and liquid streams: multiphase turbulence statistics and influence of interfacial instability

## Y. Ling[1]†, D. Fuster[2], G. Tryggvason[3] and S. Zaleski[2]

[1]Department of Mechanical Engineering, Baylor University, Waco, TX 76798, USA

[2]Sorbonne Universités, UPMC Univ Paris 06, CNRS, UMR 7190, Institut Jean Le Rond d'Alembert, F-75005 Paris, France

[3]Department of Mechanical Engineering, Johns Hopkins University, Baltimore, MD 21218, USA

The two-phase mixing layer formed between parallel gas and liquid streams is an important fundamental problem in turbulent multiphase flows. The problem is relevant to many industrial applications and natural phenomena, such as air-blast atomizers in fuel injection systems and breaking waves in the ocean. The velocity difference between the gas and liquid streams triggers an interfacial instability which can be convective or absolute depending on the stream properties and injection parameters. In the present study, a direct numerical simulation of a two-phase gas-liquid mixing layer that lie in the absolute instability regime is conducted. A dominant frequency is observed in the simulation and the numerical result agrees well with the prediction from viscous stability theory. As the interfacial wave plays a critical role in turbulence transition and development, the temporal evolution of turbulent fluctuations (such as the enstrophy) also exhibits a similar frequency. In order to investigate the statistical response of the multiphase turbulence flow, the simulation has been run for a long physical time so that time-averaging can be performed to yield the statistically converged results for Reynolds stresses and the turbulent kinetic energy (TKE) budget. An extensive mesh refinement study using from 8 million to about 4 billions cells has been carried out. The turbulent dissipation is shown to be highly demanding on mesh resolution compared to other terms in TKE budget. The results obtained with the finest mesh are shown to be not far from converged results of turbulent dissipation which allow us to obtain estimations of the Kolmogorov and Hinze scales. The estimated Kolmogorov scale is found to be similar to the cell size of the finest mesh used here. The computed Hinze scale is significantly larger than the size of droplets observed and does not seem to be a relevant length scale to describe the smallest size of droplets formed in atomization.

## 1. Introduction

Mixing layers formed between parallel gas and liquid streams are commonly seen in nature and industrial applications, e.g. , breaking ocean waves and injection of liquid fuels in engines. Typically a velocity difference exists between the two streams, which triggers a shear instability on the gas-liquid interface. The interfacial instability grows and eventually causes the bulk liquid to break into small droplets, forming a two-phase mixing layer between the two streams. At the continuum level, the gas and liquid streams are immiscible, so the "mixing" layer here indeed refers to a layer consisting of a mixture of gas and a dispersion of droplets generated from the bulk liquid disintegration. The process where the bulk liquid stream breaks into a large number of small droplets is often referred to as "atomization" and the resulting gas-droplets mixture as a spray. Since the breakup of the liquid stream can be significantly enhanced by the parallel fast gas stream, this co-flowing configuration (also known as air-blast atomization) is widely used in fuel injectors (Lefebvre & McDonell 2017).

† Email address for correspondence: stanley_ling@baylor.edu

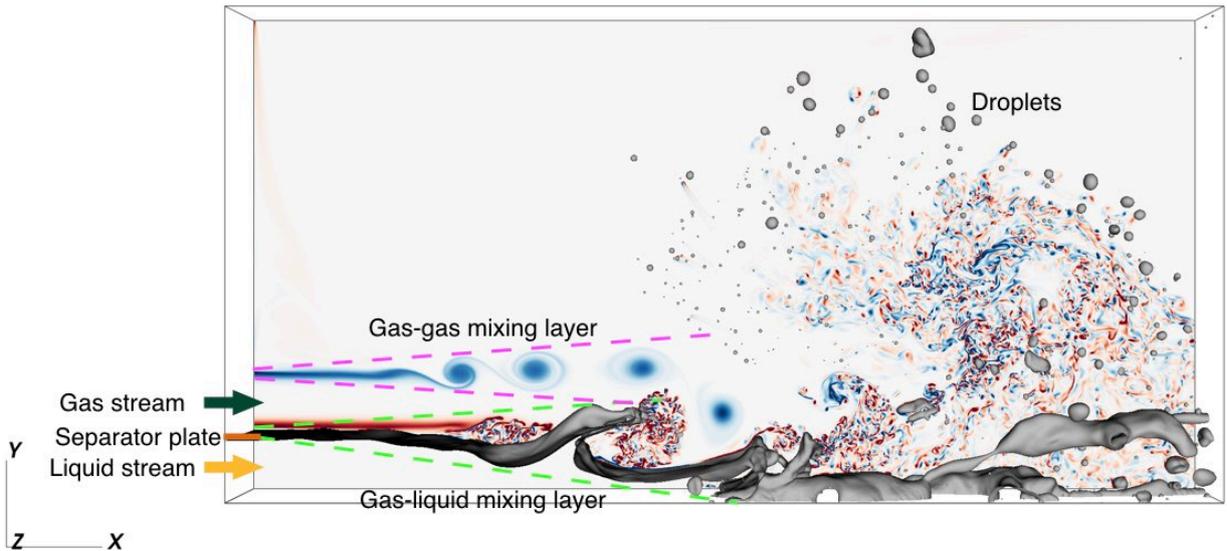

FIGURE 1. A two-phase mixing layer between parallel gas and liquid streams. The grey surface is the liquid-gas interface and the background is the z-component of vorticity.

### 1.1. *Problem description*

In the present study we focus on modeling the wall-bounded two-phase mixing layer experiment by Matas *et al.* (2011), which is illustrated in figure 1 by a snapshot of present simulation results (details of simulation are to be presented later). The grey surface is the liquid-gas interface and the background is the z-component (in spanwise direction) of vorticity. The parallel gas and liquid streams, separated by a small separator plate, enter the domain from the left. The thicknesses of the two streams at the inlet are the same. A mixing layer is formed between the gas stream and the stagnant gas, as indicated by purple dashed lines. Similarly, a two-phase mixing layer is formed between the gas and liquid streams, between the two green dashed lines. The two-phase mixing layer near the inlet is nothing but a downstream propagating interfacial wave, strictly speaking there is no "mixing". The mixing between the two immiscible phases only occur further downstream when the liquid stream breaks up, forming a mixture of gas and droplets.

As indicated in this figure, the two-phase mixing layer is a phenomenon of enormous complexity that involves interfacial instability, two-phase turbulence, and topological changes due to liquid breakups occurring at a wide variety of spatial and temporal scales. Due to its important application to fuel injection, the problem has attracted increasing attention in recent years and extensive theoretical, experimental, and numerical investi-gations have be performed. Some of the important previous works in the literature are discussed in the next sections.

## 1.2. *Shear-induced interfacial instability*

The destabilization of the liquid stream in the present problem is initiated by the instability at the gas-liquid interface, which is in turn induced by the velocity difference between parallel streams and the resulting shear on the interface near the nozzle (Taylor 1963; Renardy 1985; Rangel & Sirignano 1988; Lasheras *et al.* 1998; Lasheras & Hopfinger 2000; Matas *et al.* 2011; Matas 2015). Different mechanisms that drive the interfacial instability, including the classic Kelvin-Helmholtz instability (Helmholtz 1868; Thomson 1871), the instability due to viscosity contrast (Yih 1967), the inviscid Rayleigh (inflection-point) and viscous Tollmien-Schlichting mechanisms, have been addressed in previous works (Ozgen *et al.* 1998; Otto *et al.* 2013). Stability analysis at the interface between two immiscible fluids of different densities and viscosities has been conducted for both planar (Renardy 1985; Rangel & Sirignano 1988, 1991; Matas *et al.* 2011) and cylindrical geometries (Raynal 1997; Lasheras *et al.* 1998; Marmottant & Villermaux 2004).

Conventionally, the development of the interfacial wave is investigated through a linear analysis of small perturbations of the two-dimensional base flow (with no variation in the transverse direction) and studies focus on predicting the most unstable wavelength and frequency. The linear instability studies that yield theoretical prediction of the most unstable frequency were first carried out assuming inviscid flows (Marmottant & Villermaux 2004; Eggers & Villermaux 2008; Matas *et al.* 2011), and extensions to viscous regime have been made in recent years (Boeck & Zaleski 2005; Sahu *et al.* 2007; Fuster *et al.* 2013; Otto *et al.* 2013; O'Naraigh *et al.* 2013, 2014; Matas *et al.* 2015). Research efforts have also been made to investigate the effect of confinement on the instability of mixing layers (Juniper & Candel 2003; Juniper 2006; Juniper *et al.* 2011; Matas 2015).

While the inviscid stability theory has been shown to well predict the scaling relation of the most unstable wavelength and frequencies (Rangel & Sirignano 1988, 1991; Raynal 1997; Marmottant & Villermaux 2004; Eggers & Villermaux 2008), the viscous stability analysis is required in general to yield accurate prediction of the magnitudes of the most unstable frequency and wavelength (Fuster *et al.* 2013; Otto *et al.* 2013; O'Naraigh *et al.* 2013; Matas *et al.* 2015). The linear stability analysis in both inviscid and viscous regime confirms that the vorticity layer thickness of the gas stream at the inlet, denoted by $\delta_g$, is the characteristic length scale that controls the selection of the most unstable wavelength.

It is clearly shown by previous works (Raynal 1997; Hoepffner *et al.* 2011; Ling *et al.* 2017) that the propagation speed of the interfacial wave is well predicted by the Dimotakis speed. The Dimotakis speed $U_D$ is defined as (Dimotakis 1986)

$$U_D = \frac{\sqrt{\rho_l}U_l + \sqrt{\rho_g}U_g}{\sqrt{\rho_l} + \sqrt{\rho_g}}, \tag{1.1}$$

where $\rho_l$ and $\rho_g$ are the liquid and gas densities and the subscripts $g$ and $l$ represent the gas and liquid phases, respectively. The velocities of the liquid and gas stream at the inlet are denoted by $U_l$ and $U_g$, and $U_D$ is obtained through a phenomenological approach, assuming the gas and liquid dynamic pressures are in a balance in the reference frame moving with the wave speed. With the Dimotakis speed, the frequency and wavelength of the most unstable wave can then be related to each other as $f = U_D/\lambda$.

Recently, through viscous spatial-temporal analysis, Fuster *et al.* (2013) and Otto *et al.* (2013) showed that the interfacial instability can be absolute or convective, depending mainly on the dynamic pressure ratios between the two phases, $M$, defined as

$$M = \frac{\rho_g U_g^2}{\rho_l U_l^2}.$$ (1.2)

When $M$ is large, the interfacial instability is absolute and the wave frequency predicted by stability analysis was found to agree well with experiments and simulations (Fuster *et al.* 2013; Agbaglah *et al.* 2017).

Studies of interfacial instability eventually aim to shed light on understanding the behavior of liquid breakups occurring further downstream. Nevertheless, connections between the upstream interfacial instability and the downstream turbulence and spray characteristics have not been investigated thoroughly in previous studies. A possible reason is that a three-dimensional simulation that can resolve both the interfacial instability and the resulting turbulent spray is too expensive. In this study we only consider one specific case of two-phase mixing layer which clearly lies in the absolute instability regime. As a result, there is a dominant interfacial stability frequency and we will compare the numerical results with the spatial viscous stability theory of Otto *et al.* (2013).

### 1.3. *Two-phase turbulent coherent structures*

As the interfacial wave is formed, turbulent coherent structures simultaneously appear in the gas stream near the interface (Bernal & Roshko 1986). Due to the significant difference in velocities and viscosities between the gas and liquid streams, the gas-liquid interface acts like a deforming wavy "wall" to the gas stream. The resulting turbulent vortical structures are similar to those in boundary layers (Wu & Moin 2009; Jodai & Elsinga 2016). The growing interfacial waves significantly perturb the gas stream and play a significant role in the transition to turbulence. When the amplitude of the interfacial wave is large compared to the thickness of the gas stream, it appears as an obstacle to the gas flow, causing the latter to separate downstream of the wave. As discussed above, the frequency of wave formation will correspond to the fastest growing mode if the instability is absolute. Therefore, the turbulence production will also be related to the wave frequency. Not only can the interfacial wave development modulate the gas flow, the vortices in the gas flow also influence the wave evolution and the subsequent breakup (Jarrahbashi *et al.* 2016). The liquid stream eventually disintegrates into a large number of droplets with a wide range of sizes. These droplets are dispersed in the turbulent flow. Secondary breakup or coalescence may also occur. The present study aims to provide rigorous statistics of multiphase turbulence in the mixing layer.

### 1.4. *DNS of chaotic liquid breakups and topology changes*

Thanks to the rapid development of computer power and numerical methodology, direct numerical simulations of atomization become viable in the past decade and recent simulations have provided high-resolution details of atomization, including interfacial instability development, interaction between the interfacial wave and the turbulent gas stream, and formation of liquid sheets, ligaments, and droplets (Ménard *et al.* 2007; Shinjo & Umemura 2010; Rana & Herrmann 2011; Le Chenadec & Pitsch 2013; Jarrahbashi *et al.* 2016; Ling *et al.* 2017; Agbaglah *et al.* 2017). In particular, different droplets formation mechanisms have been observed. When the interfacial waves roll up and develop into liquid sheets, Taylor-Culick rims form at the edges of liquid sheets. Rayleigh-Taylor (RT) or Rayleigh-Plateau (RP) instabilities in the transverse direction then develop at the rims, generating liquid fingers and filaments (Marmottant & Villermaux 2004; Roisman *et al.* 2006; Agbaglah *et al.* 2013). These filaments finally break into a distribution of small droplets. In addition to this well known finger-mechanism, simulation results also reveal a less established mechanism, *i.e.*, holes form in liquid sheets and the spontaneous expansion of these holes causes liquid sheets to rupture violently, producing numerous filaments and droplets of different sizes (Shinjo & Umemura 2010; Jarrahbashi *et al.* 2016; Ling *et al.* 2017; Zandian *et al.* 2017, 2018). The holes-induced breakup of a thin liquid sheet is also observed in the bag breakup of a drop in secondary atomization (Opfer *et al.* 2014) and splashes (Marston *et al.* 2016). The mechanisms that cause sheet deformation and hole formation have been recently investigated via vortex dynamics by Zandian *et al.* (2018). In current interface-resolved simulations, disjoining pressure is generally ignored as the affordable minimum mesh size is still far larger than the sheet thickness where molecular forces are active in collapsing a liquid sheet. The holes observed in the simulations are thus an outcome of the numerical cut-off length scale, *i.e.*, the cell size (typically in microns or sub-microns). Nevertheless, recent experiments in splash and secondary breakup interestingly show that holes indeed form in liquid sheets when the thickness is around microns (Opfer *et al.* 2014; Marston *et al.* 2016). The reasons for holes arising in a thicker sheet are not fully understood, but experiments seem to indicate that the holes observed in atomization simulations are not far from what is observed in reality.

### 1.5. *Modeling of turbulent atomization*

An important future direction of atomization simulations is the development of sub-grid models like large-eddy simulation for turbulent single-phase flow (Pope 2000). Spatial scales involved in atomization processes, varying from the size of the injector to the diameter of the smallest droplet, can easily go beyond three or four orders of magnitudes. If one has to fully resolve all the scales to guarantee reasonably accurate macro-scale features, the impact of numerical simulations to practical atomization applications will be limited by their extreme costs.

Attempts to combining interface-capturing schemes and Lagrangian point-particle models have been proposed in recent years (Herrmann 2010; Tomar *et al.* 2010; Ling *et al.* 2015; Zuzio *et al.* 2017). In these combined approaches, the interfaces of the small droplets are not resolved as for the macro-scale interfaces, instead, the droplets are treated as point masses. Since the droplet-scale flows are not resolved, closure models of the force and heat transfer between the droplets and the surrounding flow are needed. As the Weber number of these droplets/bubbles are typically small, they are not much different from solid particles. Thus modeling efforts on force and heat transfer for dispersed multiphase flows or particle-laden flows are directly applicable (Magnaudet & Eames 2000; Balachandar & Eaton 2010; Ling *et al.* 2013, 2016). However, the above modeling efforts have not yet been able to resolve the fundamental challenge of sub-grid modeling of atomization, *i.e.*, how to accurately represent the under-resolved formation of sub-scale droplets. It is expected that statistics of droplets from different formation mechanisms will vary significantly. The size distribution of droplets generated in ligament breakup due to RP instability will, for example, be different from that for droplets produced in a secondary breakup.

In the literature, there are also simulations which combine interface-capturing methods and LES filtering to the turbulent gas flows (Labourasse *et al.* 2007; Larocque *et al.* 2010; Lakehal *et al.* 2012; Aniszewski 2016). These modeling efforts are mainly focused on the sub-scale surface tension effect since the small-scale variation of curvature is under-resolved. The robustness and accuracy of these models in capturing flows with significant topological changes are still to be explored. We believe that a viable sub-grid modeling approach will need to be event based. In other words, the model has to be able to identify the droplet formation event and the corresponding breakup mechanism based on topological configurations of the macro-scale liquid structures. In order to develop sub-grid model like this, rigorous data of the droplets statistics covering a sufficiently large number of events has to be collected from fully-resolved simulations.

### 1.6. *Effect of mesh resolution*

While simulations of bubbles and drops retaining their identities have been shown to produce fully converged solutions (Lu & Tryggvason 2013; Dodd & Ferrante 2016), liquid-gas multiphase flow simulations where the topology changes through breakup and coalescence generally result in spontaneously generated small-scale features that are difficult to resolve. This is particularly true for almost all simulations of large-scale atomization (Shinjo & Umemura 2010; Le Chenadec & Pitsch 2013; Jarrahbashi *et al.* 2016; Ling *et al.* 2017; Agbaglah *et al.* 2017). The general consensus among researchers has been that while the small-scale physics are under-resolved, the large-scale flow remains correct. Since small droplets and filaments contain little mass, leaving them unresolved should have only minor impact on the overall results. We have recently started to examine this assumption in more detail, by extensive grid refinement studies varying from 8 million to 4 billion cells (number of cells to resolve the initial liquid stream thickness varying from 32 to 256) (Ling *et al.* 2017). While the results show that some of the large-scale statistics converge, considerable sensitivity on the resolution has also been observed, such as for the droplet size distribution. In particular, small-scale instabilities can generate drops larger than the most-unstable wave length and the error resulting from not resolving the smallest scales fully thus manifests itself at much larger scales.

### 1.7. *Goals of study*

The purpose of the present study is to answer the following important questions for simulations of spray formation in a two-phase mixing layer between parallel gas and liquid streams:

• What are the Kolmogorov and Hinze scales in the present two-phase mixing layer and do they effectively represent the flow physics in atomization?

• What is the mesh requirement to fully resolve turbulent atomization?

• Will the large-scale multiphase turbulence statistics be affected if the small scale are under-resolved?

• How does the interfacial instability influence the multiphase turbulence development?

Particular attention will be focused on obtaining the statistics of multiphase turbulence and on the impact of the upstream interfacial instability on the turbulence. As an extension to our previous work (Ling *et al.* 2017), the simulation for the most refined mesh (M3) has been run for about twice longer time, so that the statistically converged multiphase turbulence statistics, in particular those of higher order, can be obtained.

## 2. Methodology

### 2.1. *Governing equations*

The one-fluid approach is employed to resolve the two-phase flow, where the liquid and gas phases are treated as one fluid with material properties (such as density and viscosity) that change abruptly across the interface. The incompressible two-phase flows are governed by the Navier-Stokes equations with surface tension,

$$\rho\left(\frac{\partial u_i}{\partial t} + u_j\frac{\partial u_i}{\partial x_j}\right) = -\frac{\partial p}{\partial x_i} + \frac{\partial}{\partial x_j}\left[\mu\left(\frac{\partial u_i}{\partial x_j} + \frac{\partial u_j}{\partial x_i}\right)\right] + f_{s,i}\,, \qquad (2.1)$$

$$\frac{\partial u_i}{\partial x_i} = 0\,, \qquad (2.2)$$

where $\rho$ and $\mu$ are the fluid density and viscosity, $u$ and $p$ the velocity and pressure fields. The surface tension term is expressed as

$$f_{s,i} = \sigma\kappa\delta_s n_i\,, \qquad (2.3)$$

where $\sigma$ is the surface tension coefficient (assumed to be constant here); while $\kappa$, and $n_i$ are the local curvature and unit normal of the interface. The surface tension is a singular term, with a Dirac distribution function $\delta_s$ localized on the interface.

The volume fraction $c$ is introduced to distinguish the two different phases. Here, $c = 1$ in computational cells with only the liquid phase, and its time evolution satisfies the advection equation (Hirt & Nichols 1981)

$$\frac{\partial c}{\partial t} + u_i\frac{\partial c}{\partial x_i} = 0\,. \qquad (2.4)$$

The fluid density and viscosity are calculated based on the arithmetic mean as

$$\rho = c\rho_l + (1-c)\rho_g\,, \qquad (2.5)$$

$$\mu = c\mu_l + (1-c)\mu_g\,. \qquad (2.6)$$

Detailed discussion about using arithmetic or harmonic means for viscosity has been given by Boeck *et al.* (2007). It is shown that both viscosity methods yield similar results for sufficiently high mesh resolution.

### 2.2. *Numerical methods*

The governing equations are solved by the open source code *PARIS-Simulator*. The details of the numerical methods implemented in *PARIS-Simulator* can be found in previous works (Tryggvason *et al.* 2011; Ling *et al.* 2015; Bnà *et al.* 2016; Ling *et al.* 2017) and the code webpage†. Only the numerical aspects that are relevant to the present study are summarized here.

The Navier-Stokes equations Eqs. (2.1)-(2.2), are solved by the finite volume method on a staggered grid. The fields are discretized using a fixed regular cubic grid (with cell size $\Delta$) and we use a projection method for the time stepping to incorporate the incompressibility condition (Chorin 1968). The temporal integration is done by a second-order predictor-corrector method. The interface is tracked using a volume-of-fluid (VOF) method with the mixed Youngs-centered implementation of Aulisa *et al.* (2007) to determine the normal vector and the Lagrangian-explicit scheme of Li (1995) for the VOF advection (Scardovelli & Zaleski 2003). The advection of momentum near the interface is implemented in a manner consistent with the VOF advection, similar to the methods of Rudman (1998) and Vaudor *et al.* (2017). The superbee limiter is applied in the flux calculation (Roe 1986). The viscous term is treated explicitly with a second-order centered difference scheme. Curvature is computed using the height-function method of Popinet (2009). Surface tension is computed from the curvature by a balanced continuous-surface-force method (Renardy & Renardy 2002; Francois *et al.* 2006; Popinet 2009). To capture the dynamics of under-resolved droplets less erroneously than by just quasi-fragment VOF patches, droplets of size smaller than four cells are converted to Lagrangian point-particles and are traced under the one-way coupling approximation, following the approach of Ling *et al.* (2015).

---

† The PARIS-Simulator Code, available from http://www.ida.upmc.fr/~zaleski/paris.

### 2.3. *Simulation setup*

#### 2.3.1. *Computational domain*

As shown in figure 1 the computational domain is a rectangular cuboid. The domain is initially filled with stationary gas (at $t = 0$) and then liquid and gas streams progressively enter it. The $x$-coordinate is aligned with the stream velocity; while $y$ and $z$ are along the height and width of the stream. The thicknesses of the liquid and gas streams at the inlet are represented by $H_l$ and $H_g$, respectively. Here, $H_l$ is chosen to be the characteristic length scale. Then the length $(x)$, height $(y)$, and width $(z)$ of the domain are taken to be $L_x = 16H_l$, $L_y = 8H_l$, and $L_z = 2H_l$, respectively. The thickness and the length of the separator plate are denoted as $l_y$ and $l_x$. The separator plate is included to mimic the effect of the fuel injection nozzle and the need for such a plate to accurately capture interfacial instability and wave breakups has been addressed by Fuster *et al.* (2013) and will also be discussed later.

In order to reduce the computational cost, a relatively small domain width $L_z$ is used, compared to $L_x$ and $L_y$. The characteristic length scale for the interfacial instability development is the vorticity layer thickness $\delta_g$. The current domain width is significantly larger than $\delta_g$, *i.e.*, $L_z/\delta_g = 16$, and therefore is sufficient to capture the development of interfacial stability and wave formation. When the transverse instability develops at the rim further downstream, the domain width used here may not be sufficient to resolve the large wavelengths. The effect of the domain width $L_z$ to the simulation results are discussed in the Appendix C, in which we have show results with a domain four time wider than the present one (namely $L_z = 8H_l$). The results of the present and the wider domains for both low and high order two-phase turbulence statistics (mean velocity and dissipation) agree with the results of the present domain in general, suggesting that the important conclusions made in the present study remain valid. The discrepancy mainly lies at the unbroken liquid stream near the bottom of the domain, which indicates the constraint of the domain width indeed influences the transverse instability development and interfacial wave breakup downstream in some extent. A high-resolution simulation using a wider domain is computationally expensive and will be relegated to future work.

#### 2.3.2. *Boundary conditions*

Inflow boundary condition is applied to the left of the domain $(x = 0)$, with the velocity specified as

$$
u_{x=0} = \begin{cases}
U_l \operatorname{erf} \frac{(H_l - y)}{\delta_l}, & 0 \leqslant y < H_l \,, \\
0 \,, & H_l \leqslant y < H_l + l_y \,, \\
U_g \operatorname{erf} \frac{[y - (H_l + l_y)]}{\delta_g} \operatorname{erf} \frac{[(H_l + l_y + H_g) - y]}{\delta_g} \,, & H_l + l_y \leqslant y < H_l + l_y + H_g \,, \\
0 \,, & \text{otherwise} \,.
\end{cases}
\tag{2.7}
$$

The separator plate is located at $H_l \leqslant y < H_l + l_y$. The error function, defined as

$$
\operatorname{erf}(y) = \frac{2}{\sqrt{\pi}} \int_0^y \exp(-\chi^2) d\chi \,,
\tag{2.8}
$$

is known to be the exact solution of the first Stokes problem and is employed to represent the vorticity layers on the top and bottom boundaries of the gas stream and the top of the liquid stream, following the previous works (Otto *et al.* 2013; Fuster *et al.* 2013). The thickness of the vorticity layers at the top and bottom boundaries of the gas stream is denoted by $\delta_g$ and that for the vorticity (boundary) layer at the top of the liquid stream by $\delta_l$. (There is no vorticity layer at the bottom of the liquid stream since the domain bottom is considered to be a slip wall.) For the velocity profile defined here, the displacement boundary layer thickness is $\delta/\sqrt{\pi}$ and the boundary layer thickness corresponding to $u = 0.99U_g$ is about $2\delta$ (Ling *et al.* 2017). The volume fraction function at the inlet is specified as

$$c_{x=0} = \left\{ \begin{array}{ll} 1, & 0 \leqslant y < H_l \,, \\ 0, & \text{otherwise} \,. \end{array} \right. \qquad (2.9)$$

The bottom of the domain ($y = 0$) is taken to be a slip wall and periodic boundary conditions are used at the back and front boundaries ($z = 0$ and $z = L_z$).

In order to minimize the effect of the finite size of the domain, additional attention is required for the boundary conditions at the top ($y = L_y$) and the right of the domain ($x = L_x$). In general, there are two options for the top boundary: (1) symmetric boundary (or slip wall) (Fuster *et al.* 2013; Agbaglah *et al.* 2017) and (2) free boundary that allows the gas to freely enter or leave the boundary (Taub *et al.* 2013; Ling *et al.* 2015, 2017; Almagro *et al.* 2017). If the former condition is used, a recirculating flow will form on top the parallel streams (Agbaglah *et al.* 2017). The recirculation is less favorable since obviously it may influence the physics of interest, such as carrying coherent structures downstream back to the inlet, unless the domain is so large that the effect of the recirculation becomes negligibly weak (Fuster *et al.* 2013). Due to high computational cost, a relatively small domain is used in the present study, although $L_y$ and $L_x$ are already 8 and 16 times of the initial liquid stream thickness and are large enough to capture the physics near the parallel streams. Therefore, the free boundary conditions is chosen for the top boundary in the present setup to minimize the effect of recirculation. Since we have used the free boundary condition on the top, the outlet condition on the right surface of the domain requires the convective velocity to be specified. (If a pressure outflow boundary condition is invoked, then the flow is under constrained and may exit at the top boundary, breaking the parallelism of the two streams.) The outflow velocity profile imposed at the right of the domain will affect the mean flow. In order to mimic the development of the gas stream, we specify the outflow velocity based on the average velocity of a planar turbulent jet, (Pope 2000)

$$u_{x=L_x} = \left\{ \begin{array}{ll} U_c \text{sech}^2(\alpha\xi) \,, & \text{if } y > H_l + 1/2H_g \,, \\ U_c, & \text{else} \,, \end{array} \right. \qquad (2.10)$$

where $\xi = y/y_{1/2}$, and $y_{1/2}$ is the half width of the turbulent jet. The convective velocity $U_c$ is determined by mass balance so that the flux into the domain given in (2.7) is equal to that leaving the domain. The variation $y_{1/2}$ in $x$ is found to be linear, namely, $dy_{1/2}/dx = S$. The parameters $S$ and $\alpha$ are constant, the values of which are given as 0.10 and 0.88, respectively (Pope 2000). A Neumann boundary condition $\partial c/\partial x = 0$ is applied for the volume fraction function at the right boundary.

It is noted that outflow velocity profile used here does not represent the exact condition at the outlet of the domain, since the flow is turbulent and time dependent. Equation (2.10) is thus only an approximation to the mean flow at the outlet. It is expected that the overall mean flow can be affected to a certain extent by the outflow velocity profile, in particular, a small region near the outlet will be influenced. Nevertheless, it is shown from simulation results that the overall boundary conditions applied here can effectively minimize the recirculation on the top of the parallel streams (see figure 1) and also convect the vortices and droplets out of the domain.

In order to thoroughly examine the effect of the present boundary conditions on the simulation results, we have performed simulations with a larger domain ($L_x$ and $L_y$ are 1.5 times those of the current setup) on a coarser mesh. The details of the tests for different domain sizes are given in the Appendix C. The results show that the key conclusions made in the present study are not influenced by the boundary conditions and the domain size.

### 2.3.3. *Physical parameters*

The material properties of the two fluids ($\rho_l$, $\rho_g$, $\mu_l$, $\mu_g$, $\sigma$) and the injection conditions of the two streams ($U_l$, $U_g$, $H_l$, $H_g$, $\delta_l$, $\delta_g$), values given in Table 1, fully characterize the resulting multiphase flow. In order to simplify the analysis, we take $H_l = H_g + l_y$ and $\delta_l = \delta_g$. Following the Buckingham $\pi$ theorem, the dimensional physical parameters in Table 1 are expressed in dimensionless form as shown in Table 2.

The liquid properties used here are the same as those of water. The gas is similar but not identical to pressurized air. Instead of using exact air properties in experiments (Matas *et al.* 2011; Fuster *et al.* 2013), we consider a case of moderate density ratio (Ling *et al.* 2017) that is less expensive for numerical simulation. (As will be shown later, even for this "easier" case, we barely reach fully resolved results, and thus a DNS at the exact experiment condition will be exceedingly expensive for currently available computer power.) Therefore, the fluid properties and injection conditions here are not chosen to match any realistic fuel injection condition. A larger gas density is adopted here so that the liquid-to-gas density ratio is equal to 20. The gas viscosity is chosen here so that kinematic viscosity for the two phases are the same. The dynamic pressure ratio $M$ has been shown to be the primary parameter determining the macroscale behavior of a two-phase mixing layer (Lasheras & Hopfinger 2000) and whether the interfacial instability is absolute or convective (Otto *et al.* 2013; Fuster *et al.* 2013). In order to place the interfacial instability in the absolute instability regime, a large gas-to-liquid dynamic pressure ratio is needed, and in the present study $M$ is taken to be 20. Since the liquid-to-gas density ratio $r$ and viscosity ratio $m$ are all equal to 20, we referred to this case of atomization in a two-phase mixing layer as the "A20" case.

The gas vorticity layer thickness $\delta_g$ is the characteristic length scale for the interfacial instability (Eggers & Villermaux 2008; Matas *et al.* 2011). The vorticity layer thickness $\delta_g$ varies with gas properties and injection conditions, and thus a precise value of $\delta_g$ is generally unknown a priori. In the experiment by Fuster *et al.* (2013), the injected air is at standard condition and an empirical correlation of $\delta_g$ was given as a function of the Reynolds number of the gas stream, $\mathrm{Re}_{g,H}$, which is defined as

$$\mathrm{Re}_{g,H} = \frac{\rho_g U_g H_g}{\mu_g} \, . \tag{2.11}$$

For the present simulation, the gas properties and injection condition are different, therefore, the empirical correlation of Fuster *et al.* (2013) is not applicable. The vorticity layer thickness in the present simulation is an independent parameter and the value used in the present stimulation is chosen as $\delta_g/H_l = 1/8$ (or $\delta_g/l_y = 4$).

The effect of $\delta_g$ (and $\delta_l$) on the development of interfacial instability has been discussed extensively by Fuster *et al.* (2013) through 2D simulations. To investigate the influence of $\delta_g$ on the interfacial wave breakup and the multiphase turbulence, simulations have to be extended to 3D. A parametric study of $\delta_g$ with 3D simulations is interesting yet out of the scope of the present work.

| $\rho_l$ | $\rho_g$ | $\mu_l$ | $\mu_g$ | $\sigma$ | $U_l$ | $U_g$ | $H_l$ | $\delta_g$ | $l_y$ |
|----------|----------|---------|---------|----------|-------|-------|-------|------------|-------|
| $(kg/m^3)$ | $(kg/m^3)$ | $(Pa\,s)$ | $(Pa\,s)$ | $(N/m)$ | $(m/s)$ | $(m/s)$ | $(m)$ | $(m)$ | $(m)$ |
| 1000 | 50 | $10^{-3}$ | $5 \times 10^{-5}$ | 0.05 | 0.5 | 10 | $8 \times 10^{-4}$ | $1 \times 10^{-4}$ | $2.5 \times 10^{-5}$ |

Table 1. Physical parameters.

| $M$ | $r$ | $m$ | $\mathrm{Re}_{g,\delta}$ | $\mathrm{We}_{g,\delta}$ | $\mathrm{Re}_{g,H}$ |
|-----|-----|-----|--------------------------|--------------------------|---------------------|
| $\rho_g U_g^2/(\rho_l U_l^2)$ | $\rho_l/\rho_g$ | $\mu_l/\mu_g$ | $\rho_g U_g \delta_g/\mu_g$ | $\rho_g U_g^2 \delta_g/\sigma$ | $\rho_g U_g H_g/\mu_g$ |
| 20 | 20 | 20 | 1000 | 10 | 7750 |

Table 2. Key dimensionless parameters.

The separator plate thickness $l_y$ can have a significant impact on the the interfacial instability if it is larger than or comparable to $\delta_g$ (Fuster *et al.* 2013). When $l_y/\delta_g$ is sufficiently small, then the effect of $l_y/\delta_g$ becomes negligible. Here, we chose $l_y/\delta_g = 1/4$, which is significantly smaller than the threshold value of $l_y/\delta_g = 1$ given by Fuster *et al.* (2013) and thus the specific value of $l_y$ is immaterial to the results presented here.

The Reynolds and Weber numbers of the gas stream based on the vorticity layer thickness at the inlet $\delta_g$, namely,

$$\mathrm{Re}_{g,\delta} = \frac{\rho_g U_g \delta_g}{\mu_g}\,, \tag{2.12}$$

$$\mathrm{We}_{g,\delta} = \frac{\rho_g U_g^2 \delta_g}{\sigma}\,, \tag{2.13}$$

are also key dimensionless parameters for the interfacial instability (Otto *et al.* 2013).

### 2.3.4. *Mesh resolution and time step*

The fields are discretized using a fixed regular cubic grid (with cell size $\Delta$). Simulations are performed on four meshes referred to as M0, M1, M2, and M3 so that M$n$ has $H_l/\Delta = 32 \times 2^n$ points in the liquid stream layer $H_l$, see Table 3. The time step in the simulation for each mesh is computed based on time step restrictions for the convection term (the Courant-Friedrichs-Lewy (CFL) condition), the diffusion term, and the surface tension term,

$$\Delta t \leqslant \Delta t_{\mathrm{conv}} = \frac{\theta \Delta}{u_{\mathrm{max}}}\,, \tag{2.14}$$

$$\Delta t \leqslant \Delta t_{\mathrm{visc}} = \frac{\Delta^2}{6\nu}\,, \tag{2.15}$$

$$\Delta t \leqslant \Delta t_{\mathrm{surf}} = \sqrt{\frac{(\rho_g + \rho_l)\Delta^3}{\pi\sigma}}\,, \tag{2.16}$$

where $\theta$ is the CFL number.

For the M3 mesh, $\Delta = 3.15 \times 10^{-6}$ m, $\Delta t_{\mathrm{conv}} = 1.27 \times 10^{-7}$ s (assuming $u_{\mathrm{max}} = U_g = 10$ m/s and $\theta = 0.4$), $\Delta t_{\mathrm{visc}} = 1.63 \times 10^{-6}$, and $\Delta t_{\mathrm{surf}} = 4.52 \times 10^{-7}$ s. As can be seen here the convection time step restriction is the most demanding one and as a result dictates the time step in the simulation. The small time step given by the CFL condition is due to the large gas injection velocity. In the simulation, the CFL number $\theta$ is taken to be 0.4 in general. In order to confirm the simulation results are time-step independent, smaller $\theta$ like 0.2 has also been used and it is confirmed that the time step is sufficiently small.

| Run | $\Delta(\mu m)$ | $H_l/\Delta$ | Cells # | Cores # | Total core-hrs |
|-----|-----|-----|-----|-----|-----|
| M0 | 25 | 32 | $8.39 \times 10^6$ | 32 | $\sim 3 \times 10^3$ |
| M1 | 12.5 | 64 | $6.71 \times 10^7$ | 256 | $\sim 5 \times 10^4$ |
| M2 | 6.25 | 128 | $5.37 \times 10^8$ | 2048 | $\sim 1 \times 10^6$ |
| M3 | 3.125 | 256 | $4.29 \times 10^9$ | 16384 | $\sim 14 \times 10^6$ |

Table 3. Summary of simulation runs.

The domain is initially filled with stationary gas (at $t = 0$) and then liquid and gas streams progressively enters it. It takes a time period of about $tU_g/H_l \approx 200$ for the flow to reach a statistically steady state. The transient process has been shown in previous works (Ling *et al.* 2017). For the M0, M1, and M2 meshes, the simulations all start from $t = 0$ and end at about $tU_g/H_l = 1000$, 880, and 650, respectively. For the M3 mesh, the simulation was performed using $4.29 \times 10^9$ cells and 16,384 processors. Due to the high computational cost for the M3 simulation, the simulation starts from a checkpoint of the M2 simulation at about $tU_g/H_l = 200$, and is continued only up to about $tU_g/H_l = 450$.

The M3 simulations are split into multiple runs, which are conducted on the supercomputers CINECA-FERMI in Italy, LRZ-superMUC in Germany, and TGCC-CURIE in France. The M0, M1, and M2 simulations are all performed on TGCC-CURIE. The total simulation time for all four meshes is over $15 \times 10^6$ CPU core-hours. The results presented correspond to the M3 mesh, unless stated otherwise.

The results of grid and statistical convergence studies, namely evaluating the effects of the mesh resolution and the averaging time on the present results, are shown in the Appendices A and B.

## 3. Results

### 3.1. *General behavior*

When the two streams are injected into the domain, both of them are laminar and the gas-liquid interface is perfectly flat. As the two streams meet at the downstream end of the separator plate, the velocity difference between the two streams introduces a shear on the interface, which then triggers a Kelvin-Helmholtz instability. As a response of this shear-induced instability, an interfacial wave is formed as shown in the right column of figure 2. The shape of the wave at early stage is mainly influenced by the density ratio, as explained by Hoepffner *et al.* (2011). The wave propagates downstream with the Dimotakis velocity (Eq. (1.1)) which is in between the gas and liquid injection velocities (see the right column of figure 2). This is consistent with experiments by Raynal (1997), Hoepffner *et al.* (2011) and Jerome *et al.* (2013). The amplitude of the wave grows in time. At a certain stage the wave amplitude becomes comparable to the gas stream thickness, and the interaction between the interfacial wave and the gas stream becomes strong. The interaction causes the liquid sheet pulled from the wave crest to roll and to flap and eventually the liquid sheet breaks violently.

At the same time, instability also develops at the gas stream vorticity layer near the interface due to the shear. Due to the lower velocity of the liquid stream, the gas-liquid interface is seen as a deformable and wavy wall by the gas stream. The evolution of vortical structures near the interface is visualized by the $\lambda_2$ criterion (Jeong & Hussain

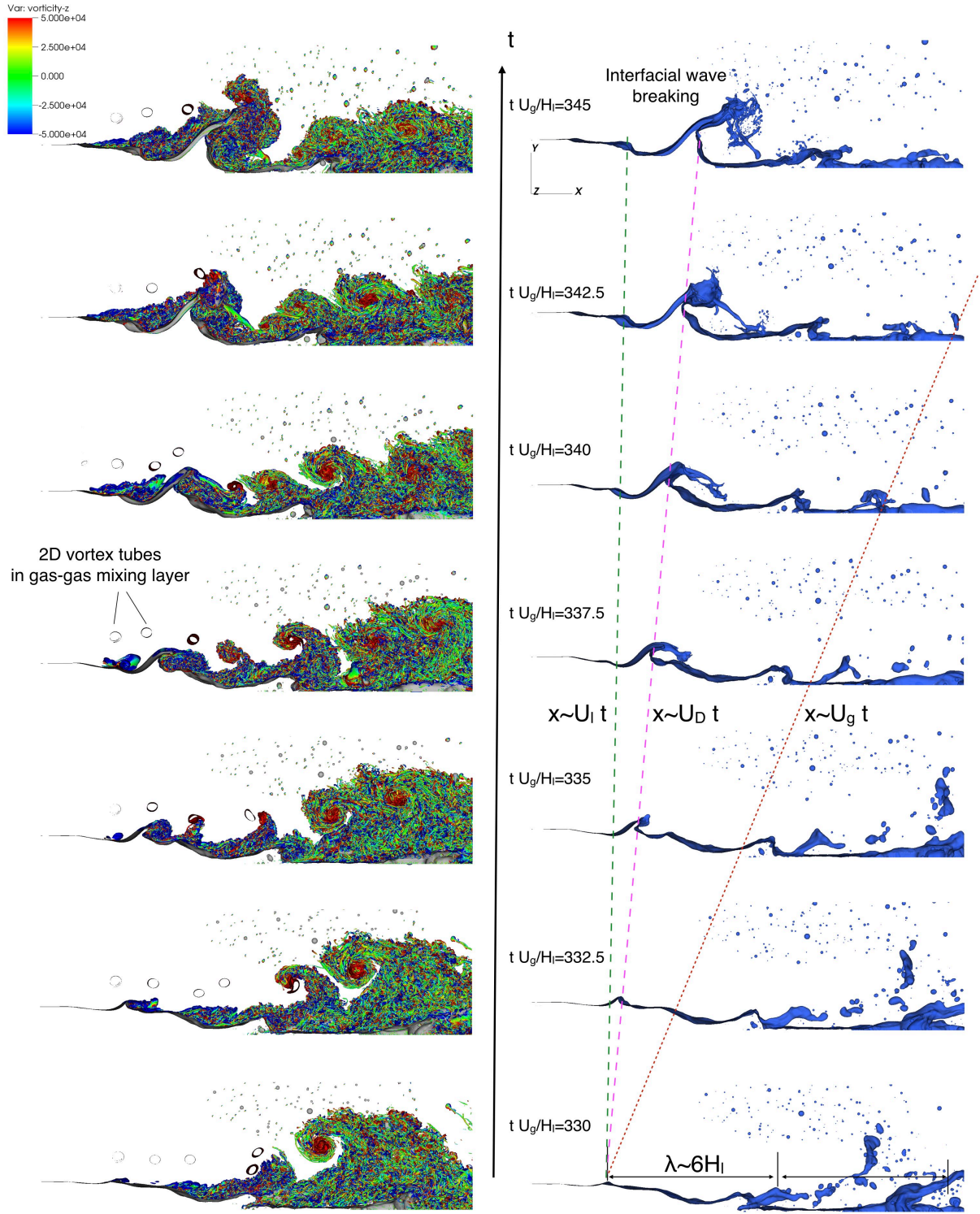

FIGURE 2. Development of the interfacial wave and coherent vortical structures.

1995). In order to distinguish the vortex rotation direction, the $\lambda_2$ iso-surface is colored by the $z-$ component of the vorticity. As a result, the red and blue vortices are aligned with the $z$ direction and rotate counter-clockwise and clockwise, respectively. On the other hand, vortices with green color are aligned with the stream direction. A 3D snapshot of the vortical structure is shown in figure 3. It can be observed that the vortical structures upstream of the interfacial wave are quite similar to those in a turbulent boundary layer (Wu & Moin 2009). The laminar vorticity layer transitions to turbulence and hairpin vortices are clearly seen near the transition region. As the amplitude of the interfacial wave becomes large and acts as an obstacle to the gas flow, the flow separates at the downstream face of the wave, forming a turbulent wake. As a result, the interfacial wave is immersed in these complex turbulent vortices and thus the stretching and breaking of the wave take place in a fully 3D chaotic manner.

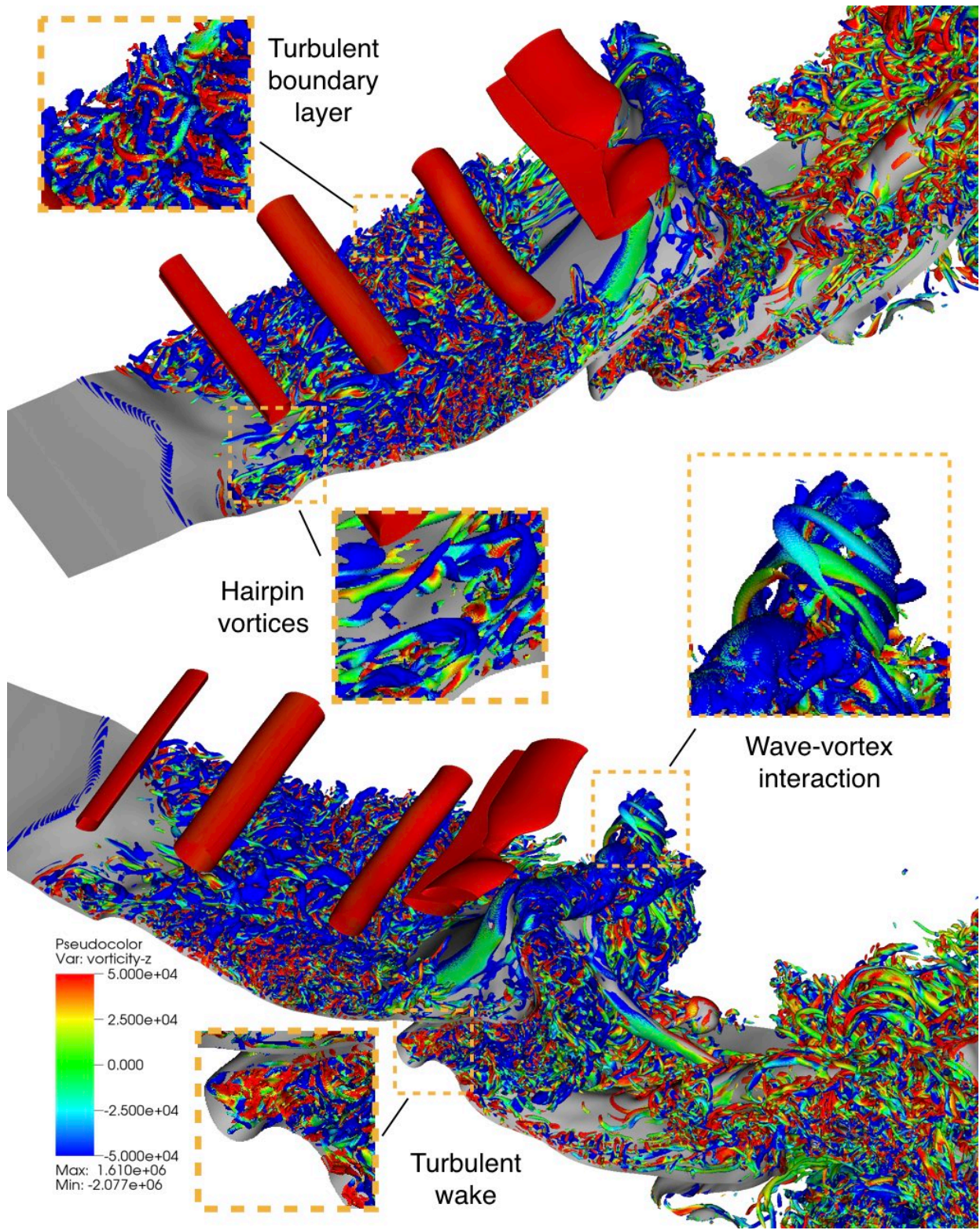

Figure 3. A snapshot of the vortical structures near the interface. Vortices are visualized by the $\lambda_2$ criterion.

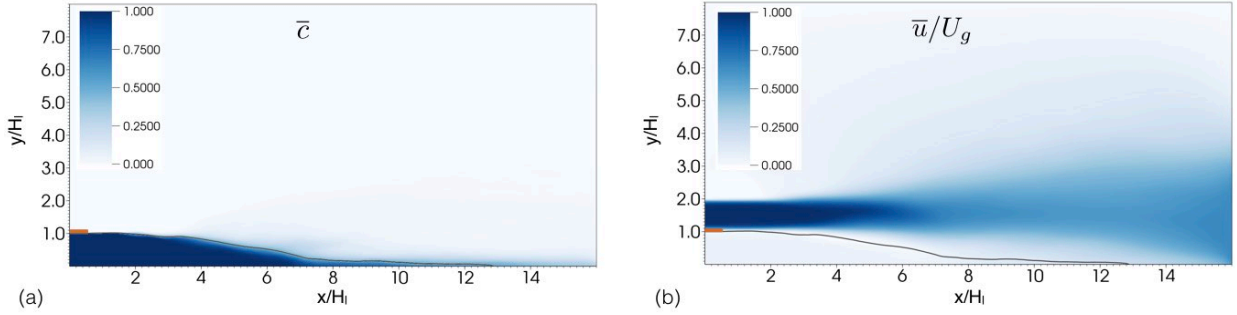

Figure 4. Average liquid volume fraction and $u$-velocity. The black curve corresponds to $\bar{c} = 0.5$. The orange rectangle near the inlet represents the separator plate.

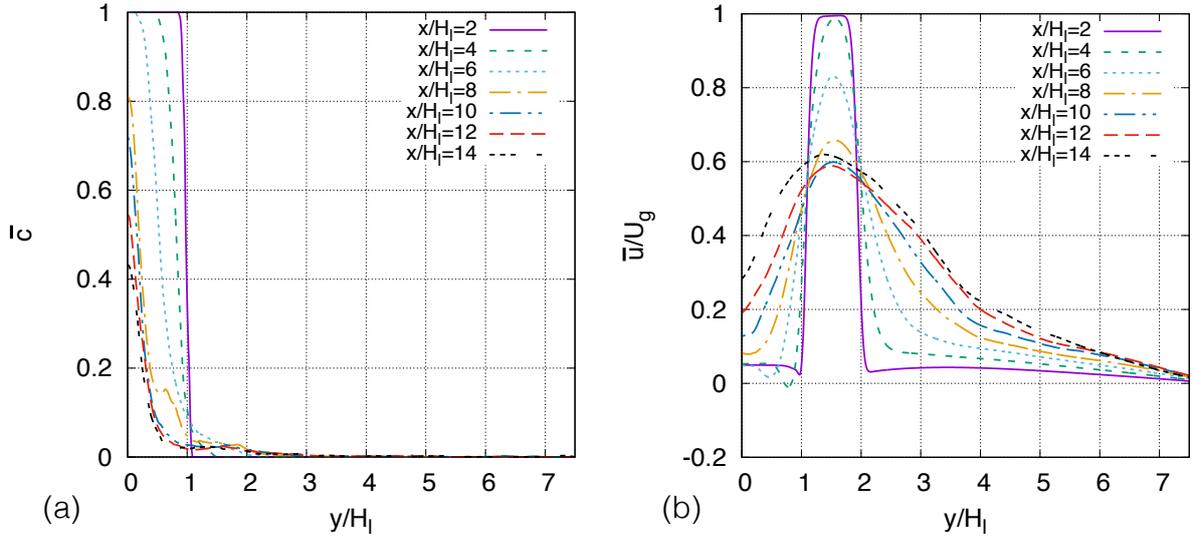

Figure 5. Mean flow profiles for Favre averaging at different streamwise locations.

### 3.2. Statistics of multiphase turbulence

#### 3.2.1. Reynolds averaging and the mean flow

The mean flow for the present problem is two dimensional ($x$-$y$), so averaging of quantities obtained from the DNS is conducted both temporally and spatially in the $z$ direction. The time and spatial (in $z$ direction) averaging operator $\overline{()}$ is defined as

$$\bar{u}(x,y) \equiv \frac{1}{t_1 - t_0} \frac{1}{L_z} \int_{t_0}^{t_1} \int_0^{L_z} u(x,y,z,t) \, dz \, dt \tag{3.1}$$

where $t_0$ and $t_1$ are the starting and ending time for averaging. In the present study, $t_0 U_g / H_l = 200$ when the statistically steady state is reached, and $t_1$ is the end time of the computation. The mean quantities are time-independent if $t_1$ is sufficiently large.

The average liquid volume fraction and the streamwise velocity are shown in figure 4. The contour line in figure 4 corresponds to $\bar{c} = 0.5$, which can be viewed as the "average" boundary of the unbroken liquid stream. The streamwise evolutions of the profiles of $\bar{c}$ and $\bar{u}$ are shown in figure 5.

The fluctuation deviating from the average quantity is given as

$$u' = u - \bar{u}, \tag{3.2}$$

which is denoted by a superscript $'$. Conventionally, the Reynolds stress tensor divided by density (or often simply referred to as Reynolds stress (Pope 2000)) is expressed as the velocity covariances,

$$\tau_{\mathrm{RA}}/\rho = \overline{u_i' u_j'}\,, \tag{3.3}$$

where the subscript RA represent Reynolds averaging. The results for $\overline{u_i' u_j'}$ are shown in figure 6. It can be observed that the maximum magnitude of $\overline{u'u'}/U_g^2$ is about 0.12, which is much larger than those of other components.

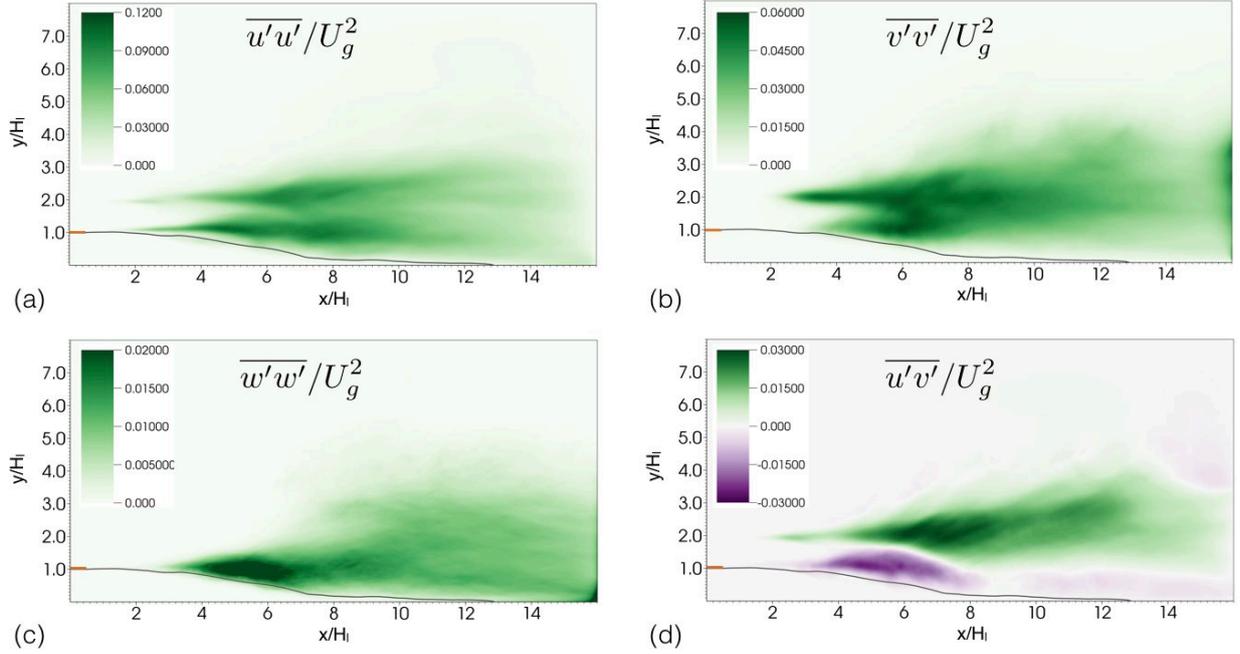

FIGURE 6. Reynolds stress tensor divided by density for Reynolds averaging. The black curve corresponds to $\bar{c} = 0.5$. The orange rectangle near the inlet represents the separator plate.

It should be note that, the Reynolds stress tensor expression given in Eq. (3.3) is strictly valid only for single-phase incompressible flows. For the present problem that involves two fluids of different density, the Reynolds stress tensor based on Favre-averaging will better characterize the turbulent two-phase flows, discussed in the following section.

### 3.2.2. Favre averaging and averaged momentum equation

For the present problem, the density at a given location may exhibit temporal fluctuations, although the density in each phase remains constant. The density fluctuations are due to the unsteady motion of the gas-liquid interface and thus are generally strong near the gas-liquid interface. As a result, the average density $\bar{\rho}$ varies spatially. For turbulent flows with variable density, such as compressible turbulent flows (Huang *et al.* 1995), the Favre-averaging (density-weighted averaging) technique is commonly employed to develop the averaged equations. The Favre averaging or decomposition have also been applied to gas-liquid flow for turbulence statistics analysis and model development (Vallet *et al.* 2001; Demoulin *et al.* 2007; Mortazavi *et al.* 2016).

The Favre-averaging operator $\tilde{(\ )}$ is defined as

$$\tilde{u} = \overline{\rho u}/\bar{\rho}\,, \tag{3.4}$$

and the fluctuation away from the Favre-averaged quantity can be expressed as

$$u'' = u - \tilde{u}\,, \tag{3.5}$$

which is denoted by a superscript $''$. It can be easily shown that $\tilde{\bar{u}} = \tilde{u}$ and $\overline{u''} = \bar{u} - \tilde{u}$.

The two-dimensional averaged momentum equation can be written as

$$\frac{\partial \overline{\rho} \tilde{u}_i \tilde{u}_j}{\partial x_j} = -\frac{\partial \overline{p}}{\partial x_i} + \frac{\partial}{\partial x_j} \overline{\left[ \mu \left( \frac{\partial u_i}{\partial x_j} + \frac{\partial u_j}{\partial x_i} \right) \right]} + \overline{f_{s,i}} + \frac{\partial \tau_{ij}}{\partial x_j} \,, \tag{3.6}$$

where $\tau_{ij}$ is the Reynolds stress tensor

$$\tau_{ij} = -\overline{\rho} \widetilde{u_i'' u_j''} \,. \tag{3.7}$$

In two-phase flows with a sharp interface, the viscosity can be expressed in terms of the Heaviside function as

$$\mu = \mu_l H + \mu_g (1 - H) \,, \tag{3.8}$$

where $H = 1$ and $0$ in liquid and gas, respectively. The volume fraction function $c$ in VOF methods is identical to $H$ in cells with either liquid or gas; while for cells with an interface, $c$ is the integral of $H$ divided by the cell volume. The viscosity computed by Eq. (2.6) is exact in cells with only liquid or gas and is a numerical approximation in cells with an interface. As a result, the average viscosity $\overline{\mu}$ can be related to $\overline{c}$ as

$$\overline{\mu} = (\mu_l - \mu_g)\overline{c} + \mu_g \,, \tag{3.9}$$

and it can be easily shown that

$$\frac{\overline{\mu u}}{\overline{\mu}} = \frac{(\mu_l/\mu_g - 1)\overline{cu} + \overline{u}}{(\mu_l/\mu_g - 1)\overline{c} + 1} \,. \tag{3.10}$$

Similarly, for the average density, we have

$$\overline{\rho} = (\rho_l - \rho_g)\overline{c} + \rho_g \,, \tag{3.11}$$

$$\frac{\overline{\rho u}}{\overline{\rho}} = \frac{(\rho_l/\rho_g - 1)\overline{cu} + \overline{u}}{(\rho_l/\rho_g - 1)\overline{c} + 1} \,. \tag{3.12}$$

In the present study, $\rho_l/\rho_g = \mu_l/\mu_g$, thus

$$\frac{\overline{\mu u}}{\overline{\mu}} = \frac{\overline{\rho u}}{\overline{\rho}} \equiv \tilde{u} \,, \tag{3.13}$$

and Eq. (3.6) can be simplified as

$$\frac{\partial \overline{\rho} \tilde{u}_i \tilde{u}_j}{\partial x_j} = -\frac{\partial \overline{p}}{\partial x_i} + \frac{\partial}{\partial x_j} \left[ \overline{\mu} \left( \frac{\widetilde{\partial u_i}}{\partial x_j} + \frac{\widetilde{\partial u_j}}{\partial x_i} \right) \right] + \overline{f_{s,i}} + \frac{\partial \tau_{ij}}{\partial x_j} \,. \tag{3.14}$$

The Reynolds stress tensor can be computed by

$$\overline{\rho} \widetilde{u_i'' u_j''} = \overline{\rho u_i u_j} - \overline{\rho} \tilde{u}_i \tilde{u}_j \,, \tag{3.15}$$

which involves third order statistics.

It can be observed from the comparison between figures 6 and 7 that, all the four components of Reynolds stress tensors from Favre averaging, $\widetilde{u_i'' u_j''}$, are quite similar to those from Reynolds averaging, $\overline{u_i' u_j'}$. The discrepancy between the Reynolds- and Favre-averaged quantities is mainly located in the gas-liquid mixing layer ($y/H_l \sim 1$), particularly in the region where the interfacial waves form and grow ($4 < x/H_l < 8$). Above the contour line of $\overline{c} = 0.5$, the magnitudes of $\widetilde{u_i'' u_j''}$ are shown to be much lower than those of $\overline{u_i' u_j'}$. The unsteady motion of the gas-liquid interface and the substantial difference in turbulence intensity in the gas and liquid streams (the liquid stream remains laminar) have a strong impact to Reynolds stresses near the interface. As in the present problem, the liquid density is significantly larger than the gas density, the mass-weighted (Favre) averaged properties (such as Reynolds stresses) are weighted toward the liquid properties. Since the velocity fluctuations in the liquid stream are much weaker than those in the gas flow, the magnitudes of $\widetilde{u_i'' u_j''}$ become smaller than $\overline{u_i' u_j'}$ in the gas-liquid mixing layer. As there is no density fluctuations in the gas-gas mixing layer in general, (except when the interfacial wave occasionally invades into the gas-gas mixing layer,) the difference between the results from Reynolds and Favre-averaging is generally small.

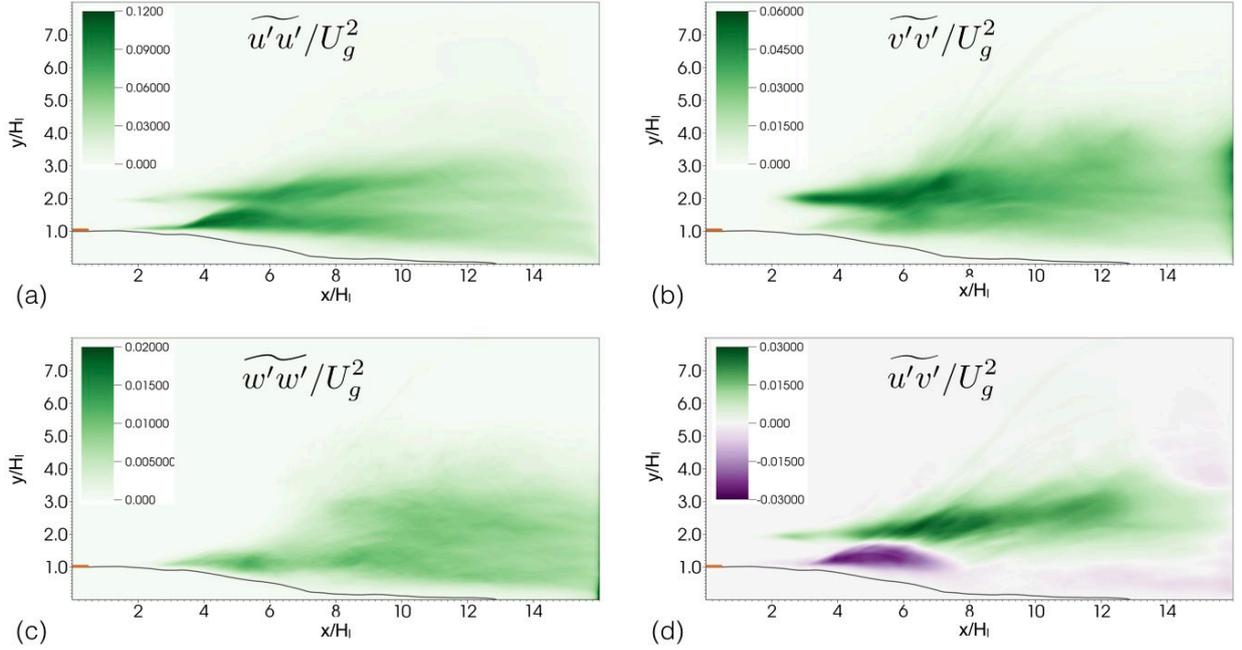

FIGURE 7. Reynolds stresses divided by the average density for Favre averaging. The black curve corresponds to $\bar{c} = 0.5$. The orange rectangle near the inlet represents the separator plate.

As shown in Eq. (3.6) that the Favre-averaging technique allows one to write the mean flow momentum equation without including density fluctuations. This is an important useful feature for two-phase turbulence modeling as already shown by Vallet *et al.* (2001). A more detailed analysis and modeling of Favre-averaged Reynolds stress tensor are of interest, yet which is out of the scope of the present work.

### 3.2.3. *Turbulent kinetic energy budget*

The equation for the kinetic energy of the instantaneous flow can be obtained by multiplying the momentum equation, Eq. (2.1), by $u_i$, giving

$$\frac{\partial \rho \frac{1}{2} u_i u_i}{\partial t} + \frac{\partial \rho u_j \frac{1}{2} u_i u_i}{\partial x_j} = -\frac{\partial p}{\partial x_i} u_i + \frac{\partial}{\partial x_j} \left[ \mu u_i \left( \frac{\partial u_i}{\partial x_j} + \frac{\partial u_j}{\partial x_i} \right) \right] - \mu \left( \frac{\partial u_i}{\partial x_j} + \frac{\partial u_j}{\partial x_i} \right) \frac{\partial u_i}{\partial x_j} + f_{s,i} u_i ,$$
(3.16)

where $u_i u_i / 2$ is the kinetic energy per unit mass. (Hereafter, we simply refer kinetic energy per unit mass as "kinetic energy" unless otherwise specified.)

Similarly, we can get the equation for the kinetic energy of the mean flow by multiplying the mean momentum equation, Eq. (3.6), by $\tilde{u}_i$ ,

$$\frac{\partial \overline{\rho} \tilde{u}_j \frac{1}{2} \tilde{u}_i \tilde{u}_i}{\partial x_j} = -\frac{\partial \overline{p}}{\partial x_i} \tilde{u}_i + \frac{\partial}{\partial x_j} \left[ \overline{\mu} \tilde{u}_i \left( \frac{\partial \tilde{u}_i}{\partial x_j} + \frac{\partial \tilde{u}_j}{\partial x_i} \right) \right]$$

$$-\overline{\mu} \left( \frac{\partial \tilde{u}_i}{\partial x_j} + \frac{\partial \tilde{u}_j}{\partial x_i} \right) \frac{\partial \tilde{u}_i}{\partial x_j} + \overline{f}_{s,i} \tilde{u}_i + \frac{\partial \tau_{ij} \tilde{u}_i}{\partial x_j} - \tau_{ij} \frac{\partial \tilde{u}_i}{\partial x_j} .$$

(3.17)

The turbulent kinetic energy (TKE), $k$ is defined as

$$k = \widetilde{u_i'' u_i''}/2$$

(3.18)

and is equal to the trace of the tensor $\widetilde{u_i'' u_j''}$, the components of which are already shown in figure 7.

The equation for TKE can be obtained by subtracting Eq. (3.17) from the averaged Eq. (3.16),

$$0 = -\frac{\partial \overline{\rho} \tilde{u}_j k}{\partial x_j} - \overline{\frac{\partial p'}{\partial x_i} u_i''} - \frac{\partial \frac{1}{2} \overline{\rho} (\widetilde{u_j'' u_i'' u_i''})}{\partial x_j} + \frac{\partial}{\partial x_j} \left[ \overline{\mu u_i'' \left( \frac{\partial u_i''}{\partial x_j} + \frac{\partial u_j''}{\partial x_i} \right)} \right]$$

$$-\overline{\mu \left( \frac{\partial u_i''}{\partial x_j} + \frac{\partial u_j''}{\partial x_i} \right) \frac{\partial u_i''}{\partial x_j}} - \tau_{ij} \frac{\partial \tilde{u}_i}{\partial x_j} + \overline{f_{s,i}' u_i''} .$$

(3.19)

Similar TKE equations have also been shown by Vallet *et al.* (2001) and Mortazavi *et al.* (2016). The terms on the right hand side are advection, pressure diffusion, turbulent diffusion, viscous diffusion, dissipation, production, and surface-tension induced diffusion, respectively. The profiles of these terms at different streamwise locations are shown in figure 8. The magnitudes of all the terms generally decrease when the sampling location moves downstream. The downstream results are more noisy, (which may be due to the fact that the averaging time is still not long enough,) but their contribution to the overall turbulence statistics is relatively small.

The TKE budget terms can be further averaged over the domain height $L_y$ to obtain a one dimensional distribution of TKE budget along the streamwise direction as shown in figure 9. Due to the existence of a large number of droplets, the term due to surface tension is generally very noisy, in particular in the downstream region where interfacial waves break into droplets, see figure 9(b). The pressure-diffusion term in Eq. (3.19) includes two contributions: the pressure fluctuations due to turbulent motion and those due to surface tension at the interface. Similar to the surface-tension term in TKE budget, the pressure diffusion also exhibits significant fluctuations (similar magnitude but with an opposite sign) downstream. Note that these fluctuations are mainly induced by the Laplace pressure at droplet interfaces instead of turbulence. To identify the pressure diffusion of TKE only related to turbulent flow motion, we plot the pressure diffusion without the contribution of surface tension (namely taking away the Laplace pressure from the total pressure), as shown in figure 9(a), the profile of which is seen to be much smoother. Furthermore, since the pressure fluctuation due to the contribution of Laplace pressure is present as long as there are interfaces. Even in the region with smooth interfaces without droplets (such as *i.e.*, $3 < x/H_l < 6$), the reduction of the magnitude of the pressure diffusion by removing the contribution of Laplace pressure is also profound.

As the TKE budget terms involve high order statistics, it is more difficult to obtain converged solutions. Results for the 1D TKE budget obtained from different meshes are shown figure 9(c). The magnitudes of the advection and the production terms are generally significantly larger than the other three terms, so three separate figures (see

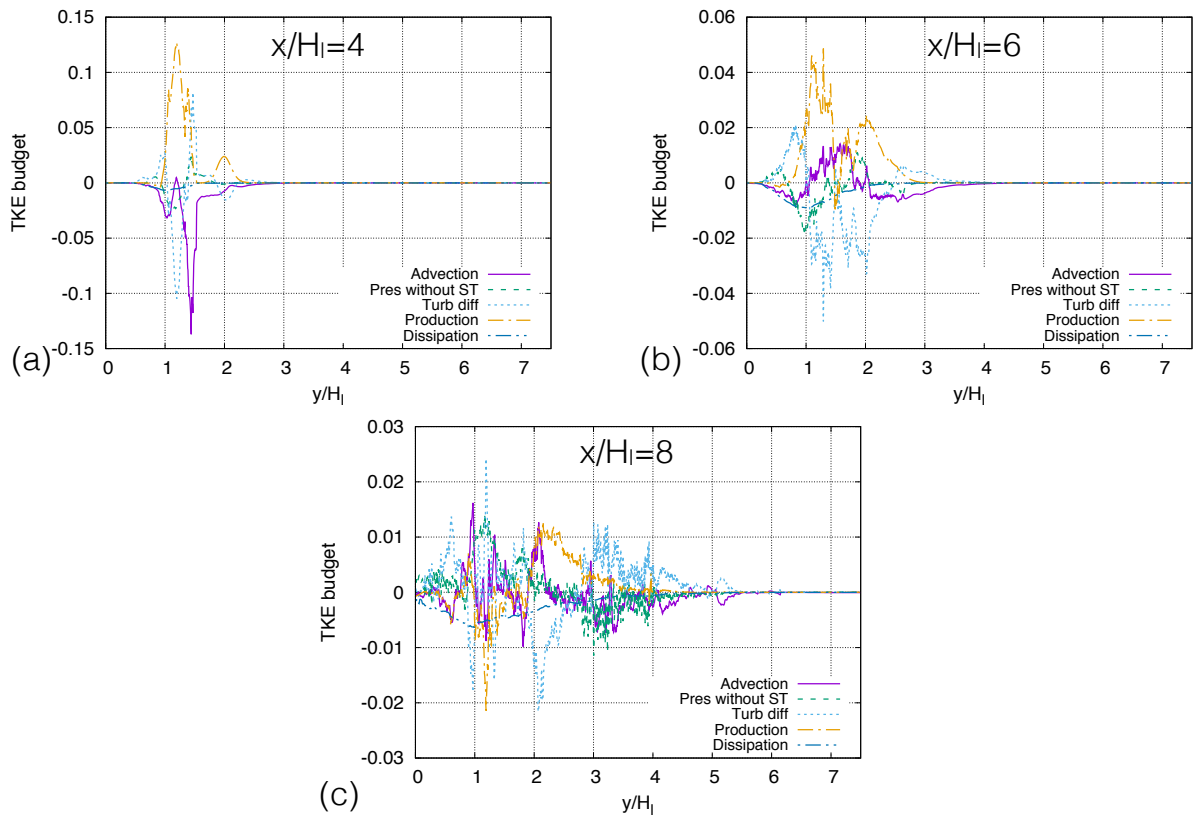

FIGURE 8. Turbulence kinetic energy budget at different streamwise locations. The terms in Eq. (3.19) are normalized by $\rho_g U_g^3 / H_l$.

figures 9(d–f)) are plotted to show the effect of mesh resolution on the dissipation, the pressure diffusion, and the turbulent diffusion terms. It can be observed that the dissipation and the pressure terms are more sensitive to the mesh resolution than other terms. In particular, minimum dissipation decreases from about -0.0005 to -0.0016 when the mesh is refined from M0 to M2. The M2 and M3 results for the dissipation agree quite well. Similar observation can be made for the pressure term. While the cell size decreases from M0 to M2, the pressure diffusion increases substantially. The results of the M2 and M3 are similar, although due to the noise in the M3 results the agreement is not as good as for the dissipation. The generally good agreement between the M2 and M3 results of high order turbulence statistics indicates that the M3 mesh is adequate to resolve the turbulence in the present problem. (Further evidence for this conclusion is to be given later based on the enstrophy calculation and the estimated Kolmogorov scale.)

It can be observed from figure 8(a) that, near the inlet at $x/H_l = 4$, TKE production in the gas-liquid mixing layer is much stronger than the gas-gas counterpart. This is consistent with previous observations in figure 2 that the two-phase mixing layer is more unstable and transits to turbulence earlier. When moving downstream, the interfacial wave grows, and vortices generated on the wave interact with the gas-gas mixing layer, accelerating its transition to turbulence. At $x/H_l = 6$, the gas-gas mixing layer produces TKE comparable to the gas-liquid counterpart. The TKE for both mixing layers are diffused effectively and at $x/H_l = 8$, the two mixing layers merge together although the results are somehow noisy. A smoother representation of the TKE budget at the downstream region would require a long averaging time, which in turn can be achieved by running the simulation for a much longer time. A longer simulation with M3 mesh is beyond the current resources available to us and will be relegated to future works.

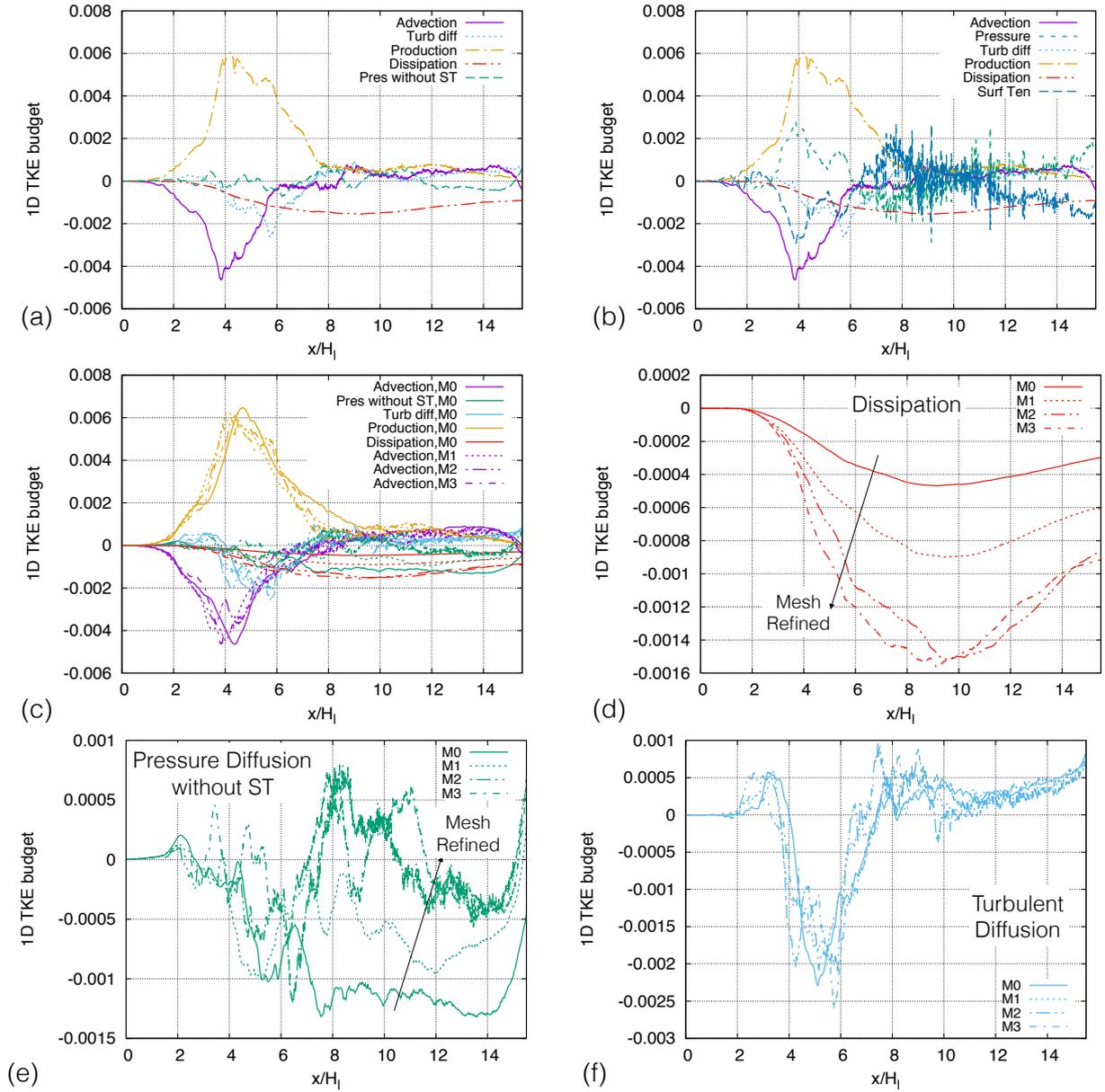

FIGURE 9. One-dimensional turbulence kinetic energy budget along streamwise direction. The terms in Eq. (3.19) are normalized by $\rho_g U_g^3 / H_l$ and averaged over domain height $L_y$. Figures (a) and (b) are TKE budget with and without the Laplace pressure contribution subtracted from the pressure diffusion term; while figure (c) shows the TKE budget computed with different mesh resolutions.

### 3.2.4. *Turbulence dissipation*

The distribution of the turbulence dissipation, denoted as $\epsilon$, for the different meshes is shown in figure 10. The results for different mesh resolutions are plotted with the same legend. (Further results of grid refinement studies can be found in the appendix, see figure 19(d).) It is clear that a fine mesh is required to capture the dissipation. While the M0 and M1 meshes underpredict turbulence dissipation, the M2 and M3 meshes yield similar results. The turbulence dissipation is generally located at the gas-liquid mixing layer. In the region of where the dissipation is larger, ($i.e.$, $4 < x/H_l < 6$), there remains a small discrepancy between the M2 and M3 results. More results for grid-refinement studies are shown in figure 19 in Appendix A. The difference between the M2 and M3 results are obviously much smaller than that between the M1 and M2 results. Therefore, although it would require a mesh finer than M3 (such as M4) to fully confirm grid independence, we believe the M3 results of dissipation presented in figure 10(a) are not far from the grid-converged solution.

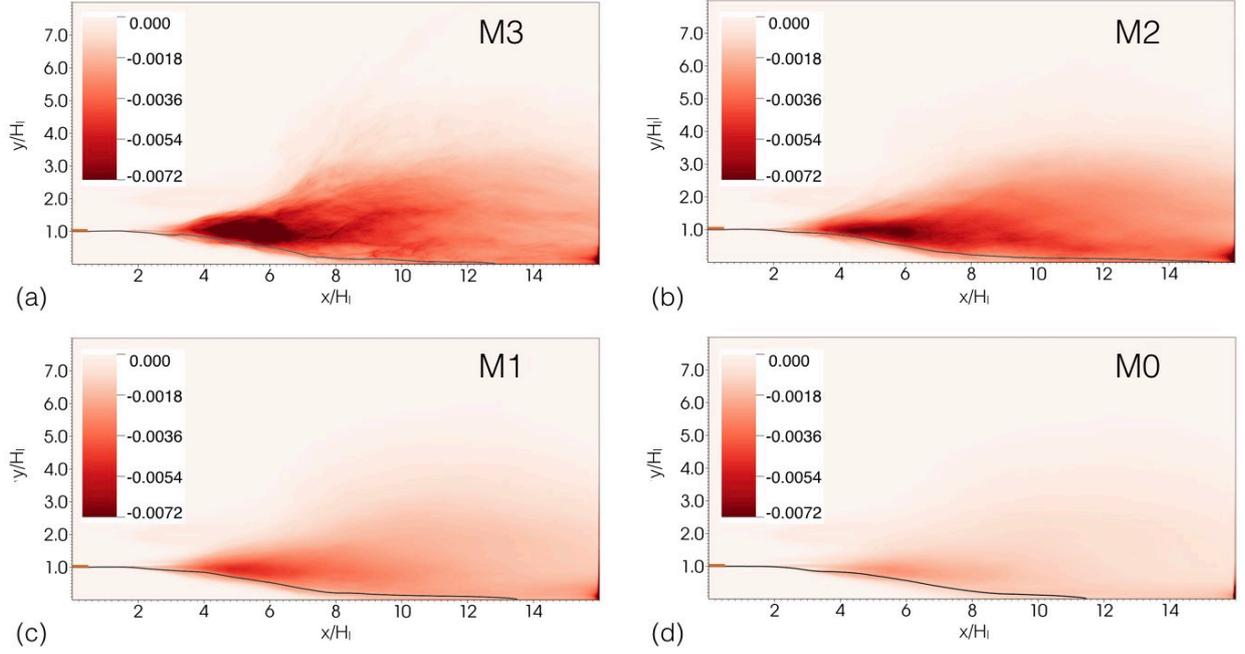

FIGURE 10. Distribution of turbulent kinetic energy dissipation for different mesh resolutions. The black curve corresponds to $\bar{c} = 0.5$. The orange rectangle near the inlet represents the separator plate.

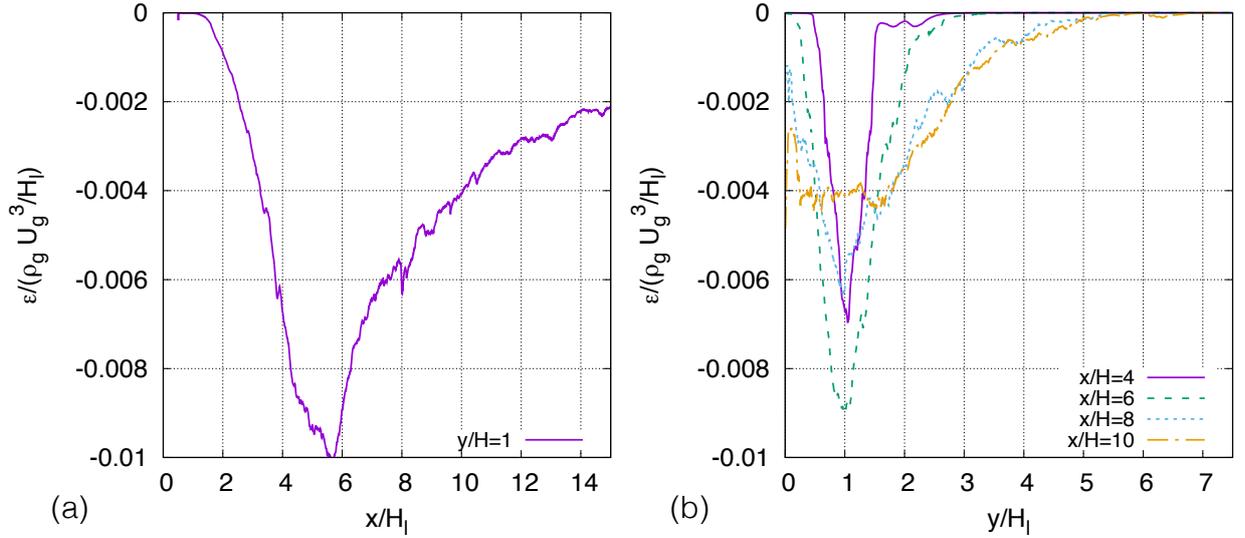

FIGURE 11. Profiles of turbulent kinetic energy dissipation along the lines (a) $y/H_l = 1$ and (b) $x/H_l = 4$, 6, 8, and 10.

The profiles of $\epsilon$ along the lines $y/H_l = 1$ and $x/H_l = 4$, 6, and 8 are plotted in figure 11. As shown in figure 11(a) the magnitude of turbulence dissipation starts to increase at about $x/H_l = 2$ where the interfacial wave starts to develop and the laminar vorticity layer transits to turbulence. The dissipation grows along $x$ as turbulence develops and reaches a maximum of about $\epsilon_{max}/(\rho_g U_g^3/H_l) = -0.01$ at about $x/H_l = 5.5$. After that, $\epsilon$ decreases gradually. Near the outlet $x/H_l = 14.5$, $\epsilon/(\rho_g U_g^3/H_l) = -0.002$. From figure 11(b) it is seen that the distribution of $\epsilon$ is initially similar to a Gaussian profile and symmetric about the line $y/H_l = 1$. As the gas-liquid mixing layer develops, the profile of $\epsilon$ expands in the $y$ direction and loses its symmetry due to the influence of the bottom wall. The bottom boundary of the non-zero $\epsilon$ region is aligned with the the contour line $\bar{c} = 0.5$. The top boundary, e.g. defined as $\epsilon = 20\%\epsilon_{\max}$, is about a straight line with the slope $dy/dx = 0.25$. This expansion with a constant slope ends at about $x/H_l = 8$, where the interfacial wave amplitude becomes comparable to the stream thickness and the two mixing layers merge. Then the distribution of $\epsilon$ becomes more uniform within the two merged mixing layer ($0 < y/H_l < 2$) as shown in both figures 10(a) and 11(b).

When details of the turbulent flow are unknown, a simple estimate of $\epsilon$ is often made based on the integral velocity $U_0$ and length scale $l_0$ as

$$\frac{|\epsilon|}{\rho_g} \approx \frac{U_0^3}{l_0}\,. \tag{3.20}$$

If we take $U_0 = U_D$ and $l_0 = H_l$, then $|\epsilon|/(\rho_g U_g^3/H_l) \approx U_D^3/U_g^3$. For the current problem with larger $M$ and $\rho_l/\rho_g$, the Dimotakis speed can be approximated as $U_D \approx U_g\sqrt{(\rho_g/\rho_l)}$, then $|\epsilon|/(\rho_g U_g^3/H_l) \approx U_D^3/U_g^3 \approx (\rho_g/\rho_l)^{3/2} = 0.011$, which is close to the maximum magnitude of dissipation obtained in simulation (see figure 11). It can be also proved that, if the gas inflow velocity $U_g$ is used as $U_0$, the $\epsilon$ will be significantly overestimated if $H_l$ remains to be used as the length scale. This seems to indicate that the interfacial wave advection speed $U_D$ is better than the inflow gas stream velocity $U_g$ in characterizing the integral scale of the turbulent flow motion.

### 3.2.5. *Estimates of Kolmogorov and Hinze scales*

With $\epsilon$ obtained above, we can estimate the Kolmogorov length scale in the gas-liquid mixing layer. The expression of the Kolmogorov length scale is given as

$$\eta = \left(\frac{\nu_g^3}{\epsilon/\rho_g}\right)^{1/4}. \tag{3.21}$$

It is shown in figure 11(a) that the maximum value of $\epsilon/(\rho_g U_g^3/H_l)$ is about 0.01. Then the corresponding Kolmogorov length scale is about 3.0 μm ($\eta/H_l = 0.0038$). In the downstream region of the gas-liquid mixing layer, $\epsilon/(\rho_g U_g^3/H_l)$ decreases to about 0.002, for which $\eta \approx 4.5$ μm ($\eta/H_l = 0.0056$). According to the DNS resolution criterion given by Pope (2000), the smallest turbulent scales will be well resolved if

$$\frac{\Delta}{\eta} \lesssim 2.1\,. \tag{3.22}$$

The cell size for the M3 mesh, $\Delta_{M3} = 3.125$ μm, clearly satisfies the criterion. Even the M2 mesh cell size is close to the required resolution. This is consistent with the observation that the M2 and M3 meshes yield similar results for dissipation (see figure 19(d)) and confirms that the M3 mesh is adequate to provide a resolved simulation of the present problem and the multiphase turbulence statistics presented above are grid-independent.

Based on a scaling argument focusing on the balance between the inertia force due to turbulent motion and the surface tension, the maximum stable droplet diameter for the droplet size was proposed by Kolmogorov (1949) and Hinze (1955) as

$$\eta_H = C\left(\frac{\sigma}{\rho_g}\right)^{3/5}\left(\frac{\epsilon}{\rho_g}\right)^{-2/5}, \tag{3.23}$$

where $C \approx 0.725$ is a constant and $\eta_H$ is often referred to as the Hinze scale. For droplets/bubbles larger than $\eta_H$, the surface tension will not be sufficient to balance the dynamic pressure fluctuations and these droplet/bubbles will break into smaller ones. Therefore, the Hinze scale indicates the smallest droplet size which can exist in a turbulent flow.

Similar to the Kolmogorov scale, we can also estimate the Hinze scale in the present problem with the turbulence dissipation obtained in simulation. For $\epsilon/(\rho_g U_g^3/H_l) = 0.01$, $\eta_H \approx 264$ μm; for $\epsilon/(\rho_g U_g^3/H_l) = 0.002$ at the downstream mixing layer, $\eta_H \approx 502$ μm.

The size distribution of droplets has been shown in our previous studies (Ling *et al.* 2017) and it was found that the majority of droplets generated in the mixing layer are significantly smaller than $\eta_H$ obtained above. (The measured mean volume-based diameter is about 50 μm, see figure 13(c) in the work by Ling *et al.* (2017).) Therefore, the Hinze scale does not well represent the size of droplets formed in the present problem.

The maximum stable droplet diameter from the Kolmogorov–Hinze theory assumes that the breakup of a large droplet is mainly dictated by the turbulent velocity fluctuation over a length comparable to the droplet diameter. Therefore, it is a good estimate of the droplet size when turbulence is responsible for breaking bulk liquids into small droplets. The disagreement between the Hinze scale and the droplet size in the present problem seems to indicate that, although the breakups of liquid sheets and ligaments are surrounded by turbulent vortices, the turbulent velocity fluctuations are not the dominant breakup mechanism and do not dictate the size of the droplets formed. During the disintegration of a ligament or a liquid sheet, droplets much smaller than the smallest wave length are observed. The satellite droplets generated from a ligament breakup, for example, are significantly smaller than the main droplets which are of the scale of the ligament diameter. If the generated small droplets were contained in a finite region and would coalesce, the size may experience a reverse cascade back to the Hinze scale. Nevertheless, in the present problem, the droplets are rapidly convected and dispersed downstream and coalescence is rarely observed. Therefore, the Hinze scale is not a relevant length scale in describing the smallest droplet size formed in the two-phase mixing layer.

### 3.2.6. *Energy spectra*

The temporal velocity spectra at different streamwise locations of the gas-liquid mixing layer are shown in figure 12. The purpose of showing the velocity spectra is to examine if the inertial subrange can be identified, which in turn can show if the turbulence at different streamwise locations is in equilibrium or not. The spectra of the $u'$ and $v'$ fluctuations do not show a clear range with the -5/3 slope. The -5/3 slope can be better discerned from the spectrum of $w'$, namely the velocity fluctuations in the homogeneous direction. The difficulty in identifying the inertial subrange is mostly likely due to the moderate Reynolds number in the present problem, for which the width of the inertial range is not very large. Further than that, the results shown in figure 12 are from temporal data for a given spatial location and are thus quite noisy, making the identification of the inertial subrange somewhat tentative.

In order to better show the velocity spectra, we plot the spatial velocity spectra in the $z$ direction at different streamwise locations in figure 13. In order to reduce the noise, the spectra are averaged in time. Then the inertial subrange with a $-5/3$ power law is more clearly revealed in all three velocity components between $k/H_l = 0$ and 1. The lower bound wavenumber is dictated by the domain width $L_z/H_l = 2$. The inertial subrange seems to be more clear and wider when the sampling location moves downstream. This seems to indicate that the turbulence at the upstream location ($x/H_l = 4$) is out of equilibrium, which in turn is due to the strong wave-turbulence interaction in the upstream region as shown in figure 2. After the wave breaks downstream, the turbulence equilibrates and the inertial subrange is better established.

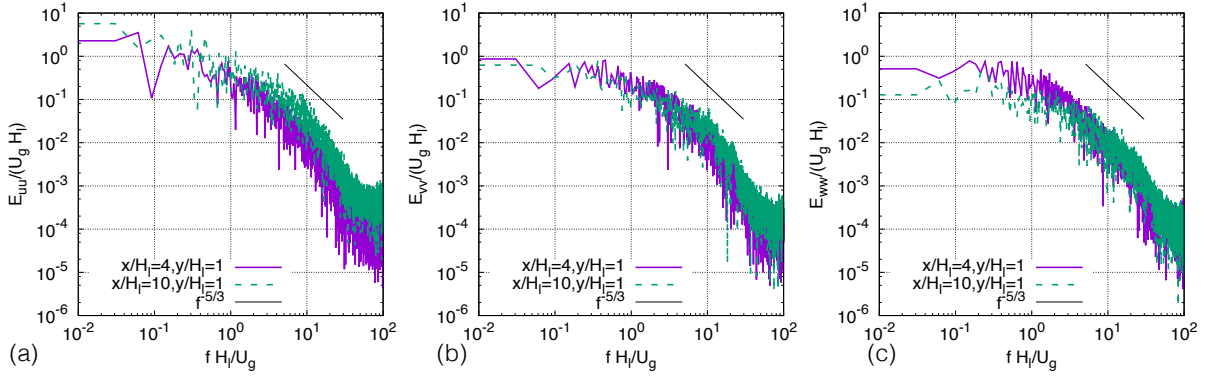

FIGURE 12. Velocity spectra at different locations.

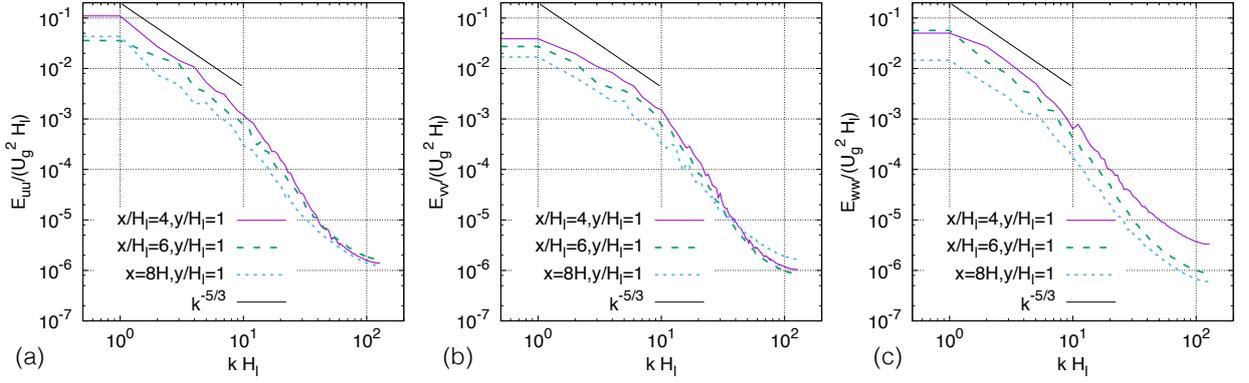

FIGURE 13. Velocity spectra at different streamwise locations.

### 3.3. *Interfacial instability regime and dominant frequency*

Following the analysis of Otto *et al.* (2013), we solve the Orr-Sommerfeld equations to investigate the viscous spatio-temporal instability of the two-phase mixing layer. The details of the approach can be found in their paper (Otto *et al.* 2013) and thus are not repeated. Here only the essential steps are briefly summarized for clarity.

The base flow is two-dimensional and the streamwise velocity profile is taken as

$$u_g(y - H_l) = -U_l \mathrm{erf}\left(\frac{y - H_l}{\delta_l}\right) + U_i\left[1 + \mathrm{erf}\left(\frac{y - H_l}{\delta_d}\right)\right], \quad y < H_l, \qquad (3.24)$$

$$u_l(y - H_l) = -U_g \mathrm{erf}\left(\frac{y - H_l}{\delta_g}\right) + U_i\left[1 - \mathrm{erf}\left(\frac{y - H_l}{\delta_d}\right)\right], \quad y > H_l, \qquad (3.25)$$

where $U_i$ is the interface velocity obtained from continuity of shear stresses across the interface as

$$\frac{U_i}{U_a} = \frac{\delta_l/\delta_g(1 + \mathcal{M}) + m(1 - \mathcal{M})}{1 + m}\frac{\delta_d}{\delta_g}, \qquad (3.26)$$

where $\mathcal{M} = (U_g - U_l)/(U_g + U_l)$, $m = \mu_l/\mu_g$ and $U_a = (U_g + U_l)/2$. The parameter $\delta_d$ is an adjusting parameter to mimic the velocity deficit behind the separator plate. Numerical simulations and experimental data reported by Fuster *et al.* (2013) have confirmed that this velocity deficit is important to capture the correct transition from convective to absolute instability.

The model base flow profiles used in stability analysis for different $\delta_d$ are shown in figure 14 which also includes the mean streamwise velocity profile at $x/H_l = 0.75$ in the present simulation for comparison.

The perturbation about the base flow are given in the form of streamfunction $\psi$ and

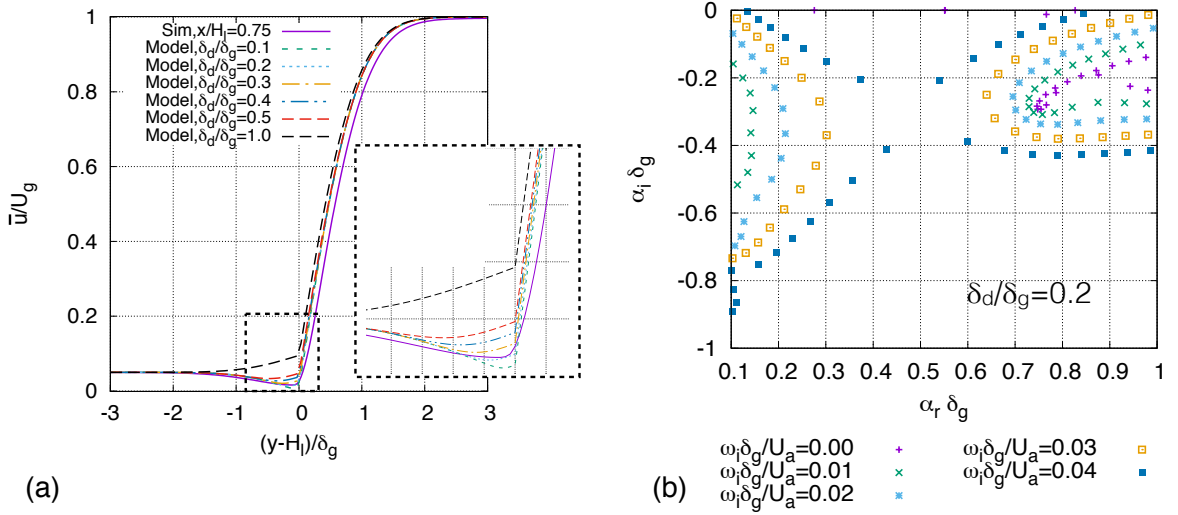

Figure 14. Spatio-temporal viscous stability. (a) Models for initial streamwise velocity profile. (b) $\omega_i$ contours to show the pinching point.

takes the form of normal modes

$$\psi_{g,l}(x, y, t) = \phi(y)_{g,l} \exp(i\alpha x - \omega t). \tag{3.27}$$

Then the Orr-Sommerfeld equations for the $\phi_g(y)$ and $\phi_l(y)$ are expressed as

$$\left[ (-i\check{\omega} + i\check{\alpha}\check{U}_l)(\frac{\partial^2}{\partial \check{y}^2} - \check{\alpha}^2) - \frac{m}{r\mathrm{Re}_{a,\delta}}(\frac{\partial^2}{\partial \check{y}^2} - \check{\alpha}^2)^2 - i\check{\alpha}\check{U}_l'' \right] \check{\phi}_l = 0, \quad y < H_l, \tag{3.28}$$

$$\left[ (-i\check{\omega} + i\check{\alpha}\check{U}_g)(\frac{\partial^2}{\partial \check{y}^2} - \check{\alpha}^2) - \frac{1}{\mathrm{Re}_{a,\delta}}(\frac{\partial^2}{\partial \check{y}^2} - \check{\alpha}^2)^2 - i\check{\alpha}\check{U}_g'' \right] \check{\phi}_g = 0, \quad y > H_l, \tag{3.29}$$

where $\check{(\;)}$ denote non-dimensional variables with $U_a$ and $\delta_g$ as typical velocity and length scales.

The branches of the imaginary part of the frequency $\omega_i$ are shown in the complex spatial wave number plane ($\alpha_r$-$\alpha_i$ plane) in figure 14(b). This diagram reveals the convective nature of the instability of the gas-liquid mixing layer. The method of Bers (1983) is used to determine the transition from convective to absolute instability. It is observed that the two branches for $\omega_i\delta_g/U_a = 0.04$ reconnect at a saddle point ($\alpha_r\delta_g = 0.5$ and $\alpha_i\delta_g = -0.3$). The value of $\omega_r$ corresponding to the saddle point is the dominant frequency emerging in the flow field. The values obtained from the theory range from $fH_l/U_g \approx 0.031$ to $0.035$ for $\delta_d/\delta_g = 0.1$ to $0.5$. Similarly, the value of $-\alpha_i\delta_g = 0.3$ is the dominant spatial growth rate. The theoretical predictions of dominant spatial growth rate and frequency are compared with simulation results in figure 15.

The mixing layer thickness here is estimated as $H_l - y_c$, where $y_c$ is the mean interfacial height corresponding to $\bar{c} = 0.5$. The spatial growth of the mixing layer thickness is shown in figure 15(a) for different meshes. It should be noted that along the streamwise direction, $H_l - y_c$ is first negative and then becomes positive (see figure 4(a)). Since $H_l - y_c$ is plotted in the log scale in figure 15(a), the results for negative $H_l - y_c$ (for $(x - l_x)/\delta_g \lesssim 9$) will not be shown. Strictly speaking, $H_l - y_c$ serves as a good measure of the mixing layer thickness only when the value is not too small, $i.e.$, $\log[(H_l - y_c)/\delta_g] > -2$. The rapid growth in the region $8 \lesssim (x - l_x)/\delta_g \lesssim 11$ is due to the artifact that $H_l - y_c$ transits from negative to positive values.

It is observed that all the M1 to M3 curves superpose for $(x - l_x)/\delta_g \gtrsim 12$. An exponential spatial growth can be identified in the region $11 < (x - l_x)/\delta_g < 16$. The exponential growth rates for the M1 to M3 meshes are quite similar. These rates agree very well with the theoretical predicted value $-\alpha_i = 0.3/\delta_g$. The exponential growth region for the M2 mesh is wider than that for the M3 mesh, which may be due to the fact that the M3 simulation has only been run for a relatively shorter time and there remain spatial fluctuations in $y_c$ (see figure 4). After that region, a more gradual nonlinear growth is seen. The linear stability theory is valid only when the amplitude of the interfacial wave and the perturbation caused by the wave to the gas stream remain small. As the interfacial wave grows and propagates downstream, the nonlinear effect will eventually become important (such as for $\log[(H_l - y_c)/\delta_g] \gtrsim -0.5$) and the simulation results will deviate from the linear theory. The fact that the exponential growth appears only in an intermediate region has also been observed in experiments (Matas *et al.* 2011) and simulations (Agbaglah *et al.* 2017).

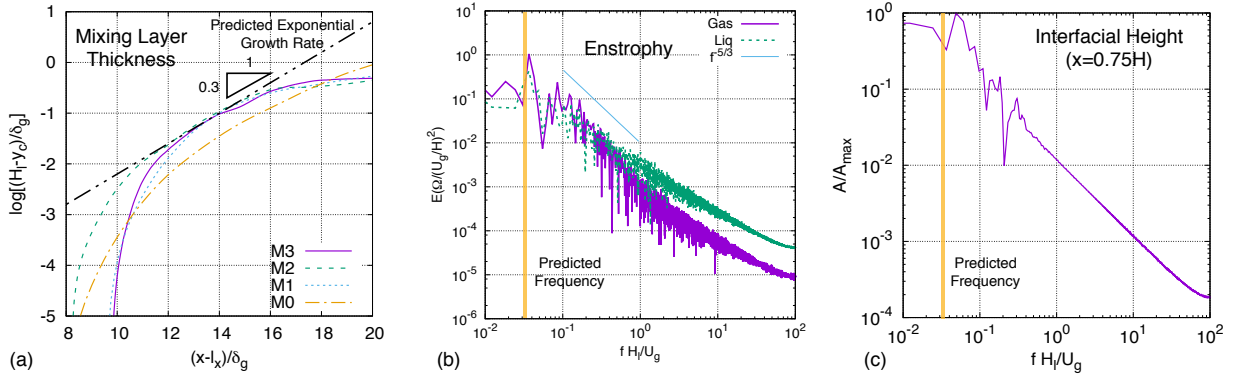

Figure 15. (a) Spatial growth of the numerically measured mixing layer thickness, compared with the exponential growth predicted using the maximum spatial growth rate; and the dominant frequency indicated by the numerical results of (b) the spectra of overall liquid and gas enstrophy of the whole domain and (c) the spectra of the interfacial height at $x/H_l = 0.75$, compared with the predicted value by stability analysis.

The dominant frequency in the simulations can be observed in the spectra of the integral of gas and liquid enstrophy over the whole domain, see figure 15(b). The liquid and gas enstrophy are computed as

$$\Omega_l = \frac{1}{2} \int_0^{L_x} \int_0^{L_y} \int_0^{L_z} c \, \omega_i \omega_i \, dx \, dy \, dz \,, \qquad (3.30)$$

$$\Omega_g = \frac{1}{2} \int_0^{L_x} \int_0^{L_y} \int_0^{L_z} (1-c)\omega_i \omega_i \, dx \, dy \, dz \,, \qquad (3.31)$$

where $\omega_i$ is the vorticity. A clear peak is seen at $fH_l/U_g \approx 0.037$ which is very close to the theoretical prediction. Figure 15(c) shows the interfacial height spectra ($A$ denotes the Fourier transform of the interfacial height) at $x/H_l = 0.75$ and the dominant frequency is about $fH_l/U_g \approx 0.05$, which is slightly larger than the theoretical prediction.

The dominant frequency (or period) can also be observed from the temporal evolution of the enstrophy as shown in figure 16. The dominant wave period is about $27H_l/U_g$ and is the inverse of the dominant frequency obtained from figure 15. Since the interfacial wave propagates with $U_D$, the wave length can be estimated as $\lambda = U_D/f \approx 6H_l$. The wavelength estimated here is consistent with the observation in figure 2. It is also observed that $\lambda$ is significantly larger than the liquid stream thickness at the inlet $H_l$.

Figure 16(a) also shows the enstrophy evolution for different mesh resolutions. The observed oscillations in the enstrophy are due to the periodic formation and breakup of interfacial waves. Both the mean value and the oscillation amplitude increase when the mesh is refined from M0 to M3, while the dominant oscillation frequencies for different mesh resolutions are similar. The coarse meshes M0 and M1 significantly underpredict the enstrophy magnitude. The average enstrophy is plotted as a function of grid size in figure 16(b). The average enstrophy increases with number of cells used to resolve the initial liquid stream thickness, *i.e.*, $H_l/\Delta$, until it saturates at the M3 mesh ($H_l/\Delta = 256$). This again indicates that the finest mesh, M3, is necessary and adequate to resolve the multiphase turbulence. We need to admit that the average enstrophy still increases about 10% from the M2 to the M3 mesh. The discrepancy is consistent to the observation in figure 10. In order to fully confirm the grid-independence of the M3 results, a simulation with an even finer mesh is required, but this will be relegated to future work.

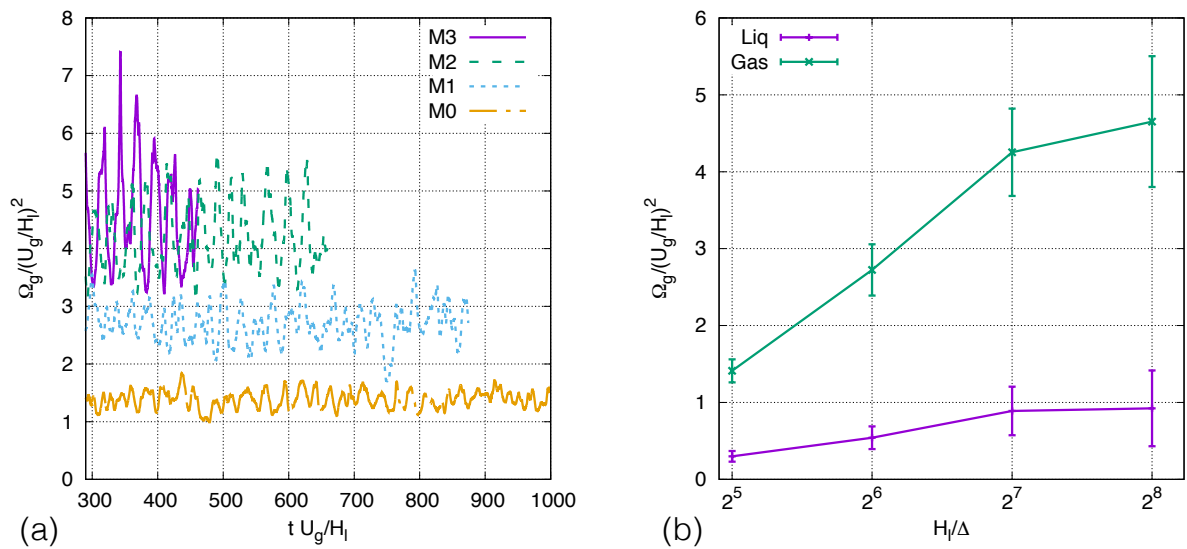

FIGURE 16. (a) Temporal evolution of gas enstrophy of the whole domain and (b) time-averaged liquid and gas enstrophy for different mesh resolutions.

The dominant frequency in the enstrophy spectra as shown in figure 15(b) is an important observation, since it clearly indicates that the turbulence production follows the same frequency as the interfacial instability. This is due to the fact that the formation and growth of the interfacial wave has a strong impact on the turbulence transition and development, see figures 2 and 3. The interfacial wave behaves as an obstacle to the gas stream and its interaction with the gas flow generates a large number of turbulent vortices both upstream and downstream of the interfacial wave. Therefore, the growth of the interfacial wave enables the kinetic energy transfer from the gas mean flow to the turbulent fluctuations. On the other hand, the flow remains laminar within the liquid stream. As a result there is a strong intermittency in the two-phase mixing layer.

In order to better illustrate the impact of interfacial wave on turbulence, we plot the temporal evolution of the root mean square (rms) of velocity fluctuations along the $z$ direction for different streamwise locations in figure 17. The averaging operator over the domain width ($L_z$) is denoted by $\hat{(\,)}$, defined as

$$\hat{u}(x,y,t) \equiv \frac{1}{L_z} \int_0^{L_z} u(x,y,z,t)\,dz \qquad (3.32)$$

and the fluctuation away from the mean value is given as

$$u^* = u - \hat{u}\,. \qquad (3.33)$$

Low-frequency fluctuations can be seen in the temporal evolution of $(\widehat{u^* u^*})^{1/2}$ measured upstream ($x/H_l = 4$), see figure 17(a), which are clearly due to the passage of the interfacial wave. The average liquid volume fraction $\hat{c}$ is equal to zero most of the time, except when the interfacial wave passes the sampling location, $\hat{c}$ jumps up to about unity, ($\hat{c} = 1$ indicates that the interfacial wave spans over $L_z$). The occurrence of spikes in $\hat{c}$ follows the period of wave formation and agree well with the dominant frequency predicted by instability theory. During the passage of the interfacial wave (within the spike), $(\widehat{u^*u^*})^{1/2}$ drops to zero; but it jumps up to a large value after the wave passes. This is due to the turbulent flows developing on the upstream side of the interfacial wave, see figure 3. Then $(\widehat{u^*u^*})^{1/2}$ will continue to decrease until the arrival of the subsequent wave.

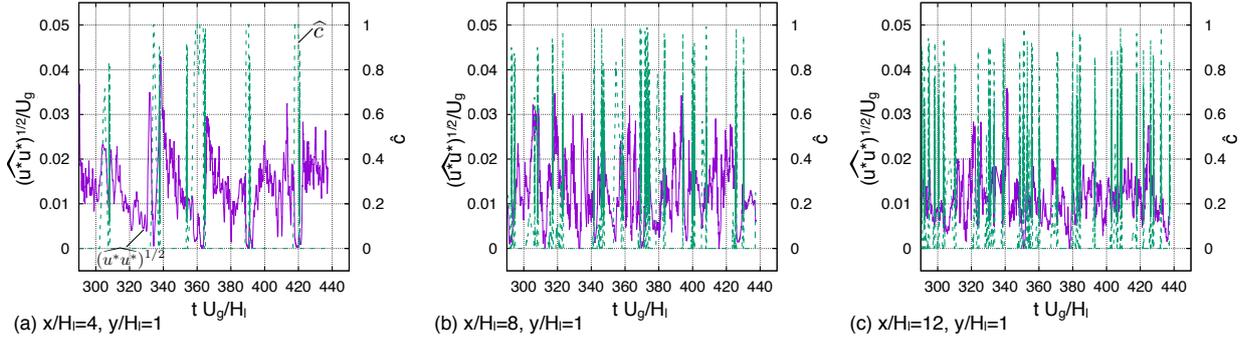

FIGURE 17. Temporal evolution of $(\widehat{u^*u^*})^{1/2}$ (purple solid lines) and $\bar{c}$ (green dashed lines) at different streamwise locations.

Further downstream ($x/H_l = 8$ and $12$), the wave breaks and the two mixing layers merge together, the effect of the interfacial instability frequency becomes less profound and the amplitude of low-frequency fluctuations in $(\widehat{u^*u^*})^{1/2}$ becomes smaller.

The temporal spectra of $\widehat{u^*u^*}$ and $\widehat{v^*v^*}$ are shown in figure 18. For $x/H_l = 4$ a dominant frequency is clearly seen at about $fH_l/U_g \approx 0.03 - 0.035$, which again agrees well with the stability theory prediction. When moving downstream at $x/H_l = 10$, the spectrum function decreases with frequency smoothly, entering the inertial regime, but no dominant frequency is observed.

Figure 18 clearly shows that the integral time scale is dictated by the dominant frequency (the most unstable mode) in the interfacial instability. The interfacial wave development is the driving force and feed energy to the resulting turbulent flows near the interface. This is also consistent with the previous observation that using the Dimotakis speed as the integral velocity scale in Eq. (3.20) better captures the dissipation.

Finally, the spectra drops at about $fH_l/U_g = 10$. With the dissipation measured in simulation, the Kolmogorov frequency can be estimated,

$$f_\eta = \left( \frac{\epsilon/\rho_g}{\nu_g} \right)^{1/2}. \tag{3.34}$$

For $\epsilon/(\rho_g U_g^3/H_l) = 0.01$, $f_\eta H_l/U_g = 8.94$, which is close to the value measured above from the spectra.

## 4. Conclusions

Direct numerical simulations of a two-phase mixing layer between parallel gas and liquid streams are performed. Particular attention is focused on obtaining high-order statistics for the multiphase turbulence arising in the two-phase mixing layer. Extensive grid refinement studies are carried out with four different meshes with the number of cells across the the liquid stream thickness varying from 32 to 256. The finest mesh (M3) consists of about 4 billion cells and is shown to be necessary and adequate to resolve the multiphase turbulence, yielding converged high-order statistics.

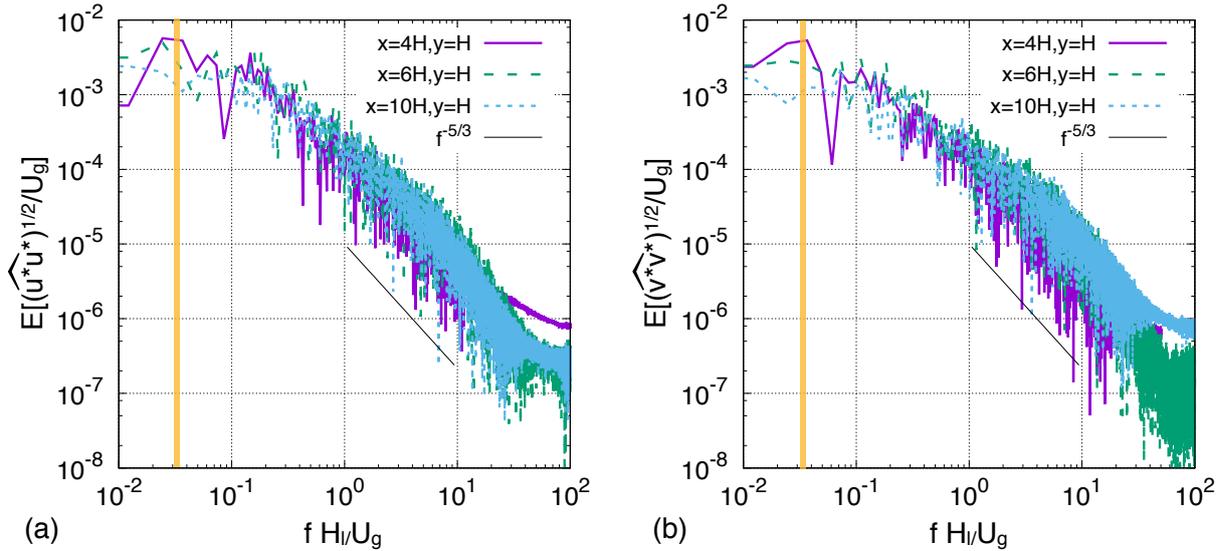

FIGURE 18. Temporal evolution of rms of fluctuations at different streamwise locations.

Due to the presence of fluids with different densities, the averaged momentum equation and the turbulent kinetic energy (TKE) transport equation are developed based on the Favre averaging technique. The results for the mean flow, Reynolds stresses, and TKE budget terms are presented. The turbulence dissipation obtained is used to estimate the Kolmogorov and Hinze scales in the present problem. The estimated Kolmogorov length scale is similar to the resolution of the finest mesh (M3) used in the present simulation, confirming that the smallest turbulent eddies are well captured. The Hinze scale is significantly larger than the typical size of droplets formed in atomization. The chaotic breakups of ligaments and sheets generate droplets that are much smaller than the most unstable wave length and the rapid droplets dispersion leave few opportunities for coalescence, therefore, the Hinze scale does not seem to well represent droplet size in primary atomization.

Viscous stability analysis is also performed on the present problem following the pervious works of Otto *et al.* (2013) and Fuster *et al.* (2013). The theory predicts that the instability of the present two-phase mixing layer is absolute, since branches on the complex wave number space reconnect at a "pinching" point. The outcome of absolute instability is that a dominant frequency will arise. The predicted value agrees well with the dominant frequency in the spectra of interface motion and also enstrophy in the domain. The dominant frequencies in interfacial instability and enstrophy are very close, which indicates that the interfacial wave development is strongly coupled with the turbulence development. Temporal evolutions of the root mean squares (rms) of velocity fluctuations for different streamwise locations are then presented. It is observed that near the inlet, there is a strong intermittence effect, *i.e.*, rms of velocity fluctuations drops to zero when a interfacial wave passes the sampling point and then rises up after the wave passage. The temporal spectra of velocity fluctuation rms exhibits a dominant frequency that match with the theoretical prediction from viscous stability analysis. The most unstable mode of interfacial instability dictates the interfacial wave period and also the integral time scale for turbulence.

| $U_D/U_g$ | $f_0 H_l/U_g$ | $\lambda/H_l$ | $|\epsilon_{\max}|/(\rho_g U_g^3/H_l)$ | $\eta/H_l$ | $f_\eta H_l/U_g$ | $\eta_H/H_l$ |
|---|---|---|---|---|---|---|
| 0.22 | 0.031-0.037 | 6.02 | 0.01 | 0.0038-0.0056 | 4.0-8.9 | 0.33-0.63 |

TABLE 4. Summary of important results for the two-phase mixing layer.

Finally the important results for the two-phase mixing layer measured in DNS are summarized in Table 4.

## Appendix A. Effect of mesh resolution

The results of grid convergence studies for the multiphase turbulence statistics are shown in figure 19. Four mesh resolutions are considered in the present study and the details are listed in Table 3. It is seen that all the four meshes used here capture the mean flow properties, including $\bar{c}$ and $\bar{u}$, very well. When it comes to higher order statistical terms, such as the Reynolds stress and the turbulence dissipation, then the M0 and M1 meshes are shown to be insufficient. It can be seen from figure 19(d) that, when the cell size decreases from the M0 to M2 meshes, the magnitude of the turbulent dissipation increases significantly. In other words, the M0 and M1 meshes significantly underpredict the dissipation. The results for both Reynolds stress and turbulence dissipation for the M2 and M3 meshes generally agree very well, indicating that the M3 mesh is adequate to resolve the multiphase turbulence in the present problem. The agreement between the M2 and M3 results for the Reynolds stress component and the dissipation at $y/H_l = 1$ and 2) is not as good as in other regions. The discrepancy is more profound for the downstream location $(x/H_l = 10)$. This is mainly due to the the fluctuations in the M3 results. The fluctuations in the M3 results can be further reduced by running the simulation for a longer time. Yet due to the extreme computational cost, the longer run can only be left for future work. In spite of the noise, the current results are sufficient to provide reasonable estimate of Reynolds stress and turbulent dissipation in the two-phase mixing layer.

## Appendix B. Effect of averaging time

The results of statistical convergence study for the multiphase turbulence statistics are shown in figure 20. Here, three different time durations are used for averaging in Eq. (3.1) where the averaging time T1, T2, T3 are 43, 77, and 135 $H_l/U_g$, respectively. The results clearly show that all the three cases well capture the mean flow properties ($\bar{c}$ and $\bar{u}$). Nevertheless, for higher order statistics like the Reynolds stress downstream and the turbulence dissipation, a longer averaging time is required. The simulation length T3 seems to yield converged results, although the turbulence dissipation at downstream location is still somehow noisy. All the results for the M3 mesh presented in the results section are averaged over T3.

## Appendix C. Effect of the domain size

In order to examine the effect of the domain size, we have also considered two different domains that are larger than the present setup: a wider domain ($L_x/H_l = 16$, $L_y/H_l = 8$, and $L_z/H_l = 8$) and a longer and higher domain ($L_x/H_l = 24$, $L_y/H_l = 12$, and $L_z/H_l = 2$). The results of the interfacial instability and the multiphase turbulence statistics are shown in figures 21 and 22, respectively. The tests are done with a mesh resolution equivalent to M1 and the results in figure 22 are plotted with the same color scale.

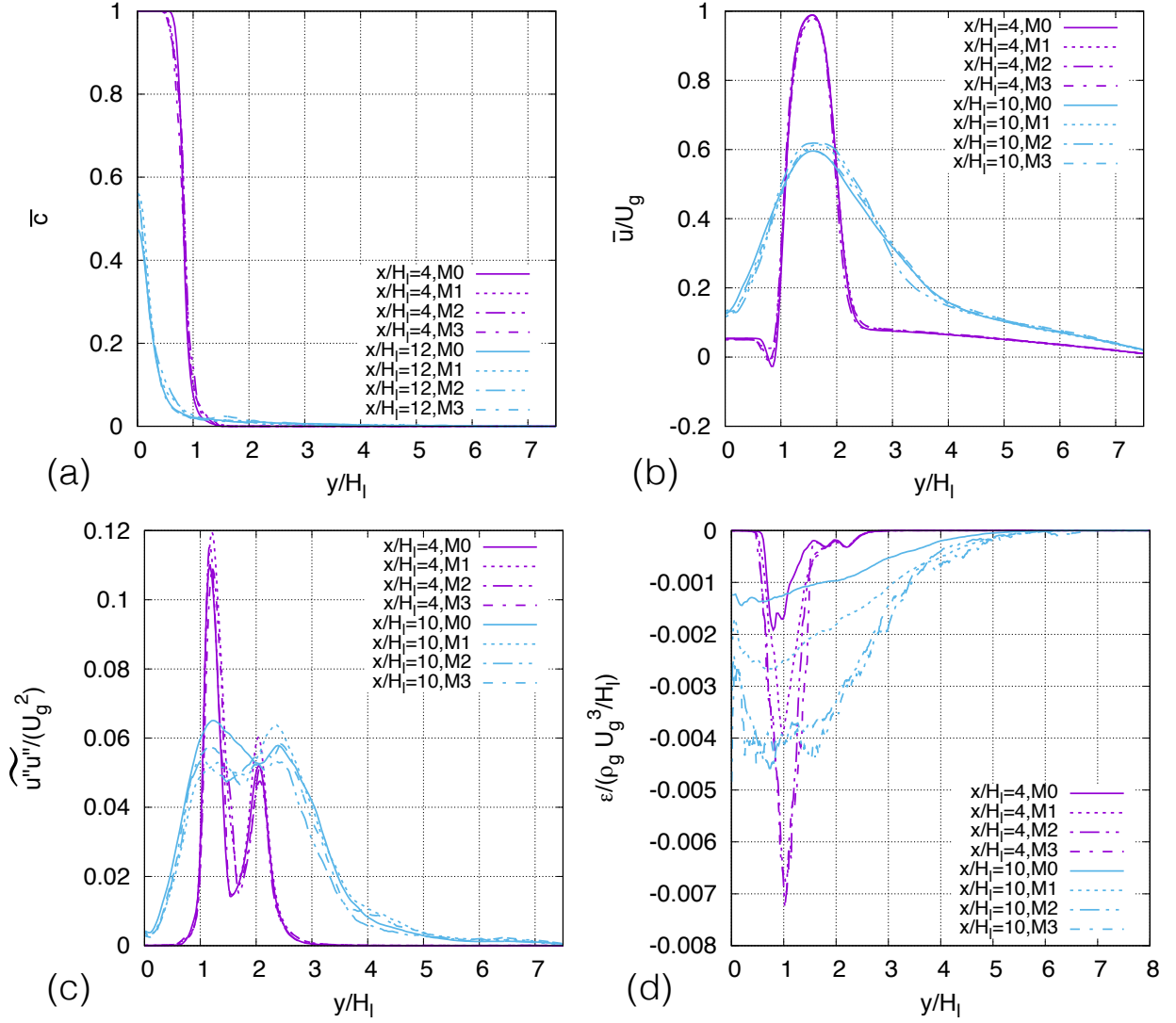

FIGURE 19. Statistical convergence studies for multiphase turbulence statistics.

The development of the interfacial instability is characterized by the vorticity layer thickness ($\delta_g$) (Matas *et al.* 2011). The width of the present domain is significantly larger than $\delta_g$ ($H_l/\delta_g = 16$) and thus is sufficient to capture the dominant frequency arising from absolute instability. Figure 21 shows the spectra of the gas enstrophy and the interfacial height for the present and the wider domains, *i.e.*, $L_z/H_l = 2$ and 8. The simulation for the wider domain was run for a shorter time, so the spectra are more noisy, nevertheless, the results for the two domains generally agree well with each other. A dominant frequency is observed in both cases, though the dominant frequency seems to shift slightly to the right for the wider domain.

As the interfacial wave grows as it propagates downstream, transverse instability develops and the wave becomes fully 3D (Zandian *et al.* 2018). Then the domain constraint in the transverse direction will influence the transverse instability (since the long wavelength instability will not be resolved) and later on wave breakup. From figures 22 (a) and (b) it can be observed that although the results of $\overline{u}$ and $\epsilon$ for the two cases generally agree well, there exists a discrepancy in the contour line $\overline{c}$, showing that the unbroken liquid stream is shorter for the case with the wider domain. This indicates that the downstream dynamics of the two-phase mixing layer would require a wider domain to avoid any influence of boundary conditions but this is left for future research.

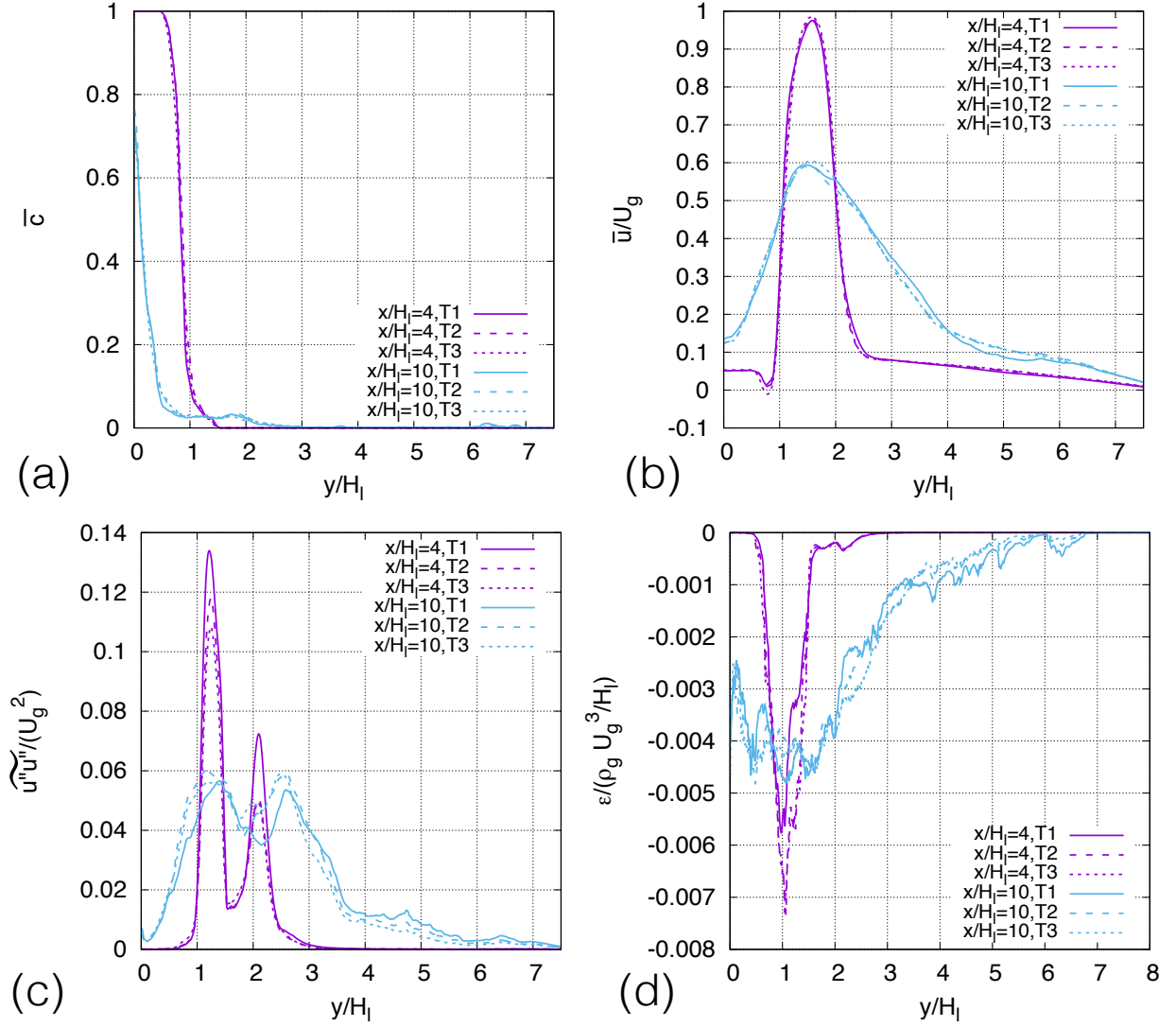

FIGURE 20. Statistical convergence studies for multiphase turbulence statistics.

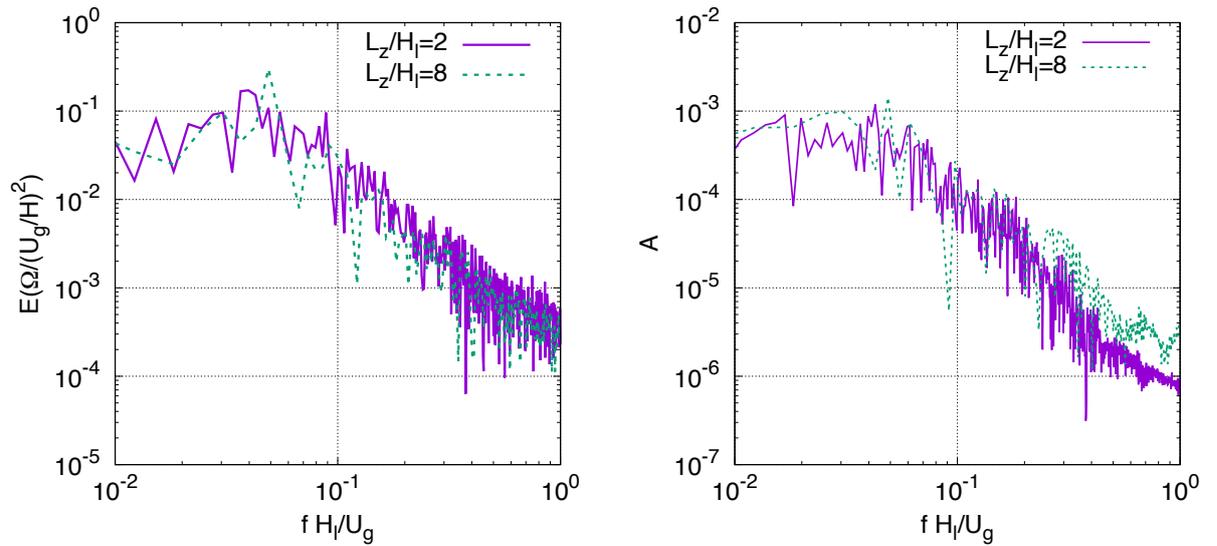

FIGURE 21. Effect of domain width on the spectra of (a) the integrated gas enstrophy in the domain and (b) the interfacial height at $x/H_l = 0.75$. $L_z/H_l = 2$ and 8 for the present and the wider domains.

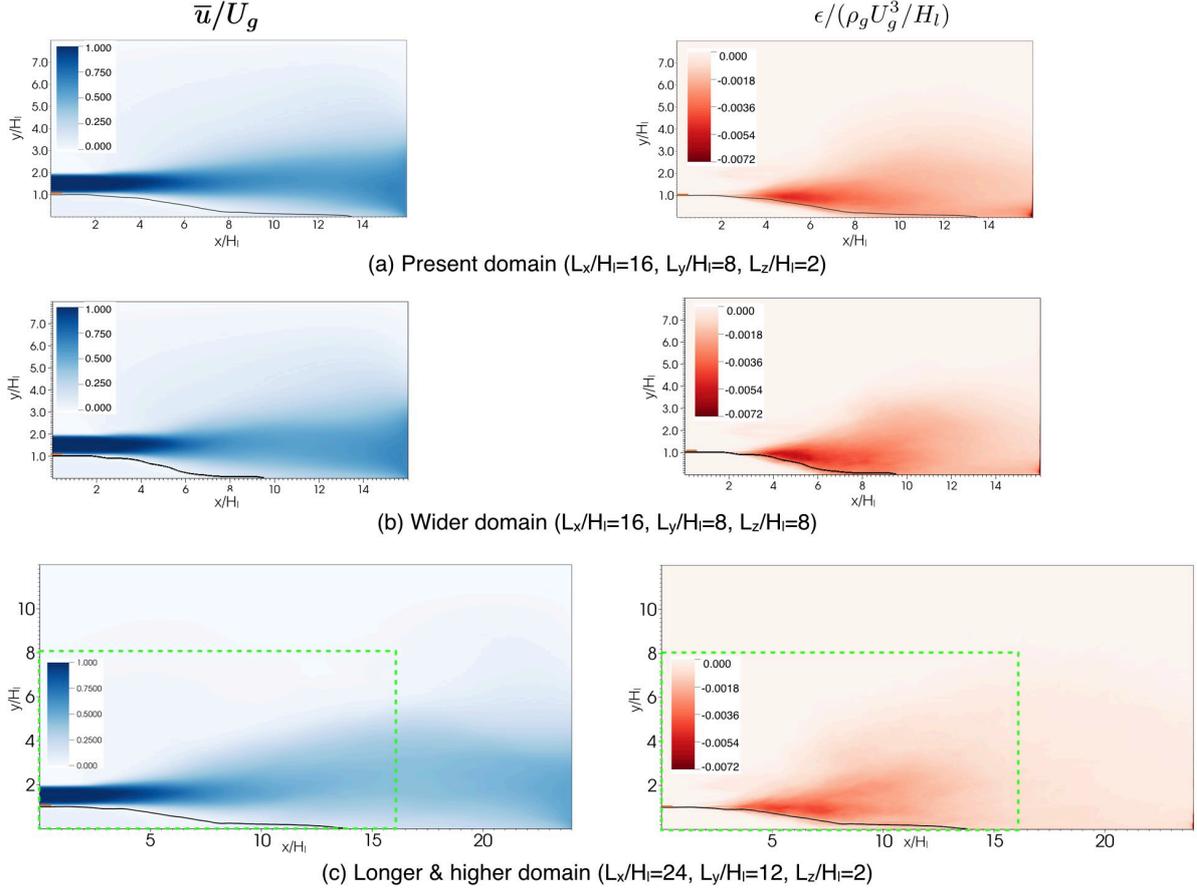

FIGURE 22. Comparison of results for different domain size: (a) the present domain ($L_x/H_l = 16$, $L_y/H_l = 8$, $L_z/H_l = 2$), (b) the wider domain ($L_x/H_l = 16$, $L_y/H_l = 8$, $L_z/H_l = 8$), and (c) the longer and higher domain ($L_x/H_l = 24$, $L_y/H_l = 12$, $L_z/H_l = 2$). The results presented here are for the M1 mesh resolution, $i.e.$, $H_l/\Delta = 64$. The left and right columns are the mean velocity and the turbulent dissipation, respectively. The black curve corresponds to $\bar{c} = 0.5$. The box with green dashed lines indicate the present domain size. The orange rectangle near the inlet represents the separator plate.

In order to avoid generating a recirculation above the parallel streams, we apply a Neumann boundary condition for the velocity on the top boundary to allow fluid to freely enter or leave the domain. Accordingly a velocity Dirichlet boundary condition is applied to the right boundary with the velocity profile specified as Eqs. (2.7) and (2.10). The outflow velocity profile given is to mimic the mean flow near the outlet and doest not represent the exact time-dependent outflow condition. The results of the mean liquid volume fraction (the black lines indicated $\bar{c} = 0.5$), the mean velocity and the turbulence dissipation for the present domain ($L_x/H_l = 16$, $L_y/H_l = 8$, and $L_z/H_l = 2$) are compared to those for a longer and higher domain ($L_x/H_l = 24$, $L_y/H_l = 12$, and $L_z/H_l = 2$). It can be seen that the results of the present domain in general agree well with those in the larger domain, except a small region near the outlet.

Therefore, the important observations made in the results section are thus confirmed to be not influenced by the domain size and the applied boundary conditions.

**Acknowledgments**

This work has been supported by the Department of Mechanical Engineering at Baylor University in United States and the ANR MODEMI project (ANR-11-MONU-0011) program in France.

This work was granted access to the HPC resources of TGCC-CURIE and CINES-Occigen under the allocations t20152b7325, t20162b7760, 2017tgcc0080, made by GENCI. HPC resources at CINECA and LRZ based in Italy and Germany have been used for the M3 mesh simulations, supported by PRACE (2014112610). This work has used resources of the Oak Ridge Leadership Computing Facility, which is a DOE Office of Science User Facility supported under Contract DE-AC05-00OR22725. The authors would also acknowledge the Texas Advanced Computing Center for providing HPC resources that have contributed to the simulation results. The Baylor High Performance and Research Computing Services (HPRCS) have been used to process the simulation results reported in this paper.

We would thank Dr. R. Scardovelli, Dr. W. Aniszewski, Dr. J. Lu, Dr. L. Malan for their contribution to the development of the code *PARIS-Simulator* and also thank Dr. T. Otto and Dr. T. Boeck for sharing their spatio-temporal stability code. We also appreciate the helpful discussions with Dr. A. Cartellier and Dr. J.-P. Matas.

Finally, the simulation data are visualized by the software VisIt developed by the Lawrence Livermore National Laboratory.